\documentclass[a4paper,fleqn,usenatbib]{mnras}

\usepackage{amsmath}
\usepackage{amsfonts}
\usepackage{amsbsy}
\usepackage{amssymb}
\usepackage{amsbsy}

\usepackage[T1]{fontenc}
\usepackage{ae,aecompl}

\usepackage{graphicx}
\usepackage{gensymb}
\usepackage{float}
\usepackage[utf8]{inputenc}
\usepackage{tensor}
\usepackage{bm}
\usepackage{enumerate}

\title[]{Gravito-turbulence and the excitation of small-scale parametric instability in astrophysical discs}

\author[]{
A. Riols,$^{1}$ H. Latter $^{1}$, S.-J. Paardekooper $^{1,\, 2}$ 
\\
$^{1}$Department of Applied Mathematics and Theoretical Physics, University of Cambridge, Centre for Mathematical Sciences, \\
Wilberforce Road, Cambridge CB3 0WA, UK. \\
$^{2}$ Astronomy Unit, School of Physics and Astronomy, Queen Mary, University of London, Mile End Road, London E1 4NS, UK}

\date{Accepted XXX. Received YYY; in original form ZZZ}

\pubyear{2017}

\begin{document}
\label{firstpage}
\pagerange{\pageref{firstpage}--\pageref{lastpage}}
\maketitle

\begin{abstract}
Young protoplanetary discs and the outer radii of active galactic
nucleii may be subject to gravitational instability and, as a
consequence, fall into a
`gravitoturbulent' state. While in this state, appreciable angular
momentum can be transported. Alternatively, the gas may collapse into
bound clumps, the progenitors of planets or stars.
In this paper, we numerically characterize the properties of 3D
gravitoturbulence, focussing especially on 
its dependence on numerical parameters (resolution, domain size)
and its excitation of small-scale dynamics.
Via a survey of
 vertically stratified shearing box simulations with PLUTO and
RODEO, we find (a) evidence that certain gravitoturbulent properties
are independent of
horizontal box size only when the box is
larger than $\simeq 40 H$, where
$H$ is the height scale, (b) at high resolution, small-scale
isotropic turbulence appears off the midplane around $z\simeq 0.5
-1 H_0$, and 
(c) this small-scale dynamics
results from a parametric instability, involving the coupling of
inertial waves with a large-scale axisymmetric epicyclic mode. This
mode oscillates at a frequency close to $\Omega$ and is naturally
excited by gravito-turbulence, via a nonlinear process to be
determined. The small-scale turbulence we uncover has
potential implications for a wide range
of disc
physics, e.g.\ turbulent saturation levels, 
fragmentation, turbulent mixing, and
dust settling.
\end{abstract}

\begin{keywords}
accretion discs -- turbulence --- instabilities --
protoplanetary discs
\end{keywords}



\section{Introduction}

Gravitational instability (GI) attacks gaseous accretion disks that are
sufficiently cold and massive, as might be the case in the early
stages of a protoplanetary (PP) disk's life \citep[e.g.][]{durisen07},  
or beyond roughly 0.01 pc
in active galactic nucleii \citep[AGN, e.g.][]{shlosman87}. 
Recent observations of wide-orbit planets
and spiral structure have fueled ongoing interest in GI
within the PP disk context \citep{Fukagawa13,christiaens14,kalas08}, while  
indications of transonic turbulence from maser emission, and
the long-standing questions of disk truncation, star formation, and
the sustenance of high accretion rates motivate its study in AGN
\citep[e.g][]{wallin98,Goodman03}. 

 The critical parameter that sets the onset of the
 gravitational instability is the Toomre $Q$ \citep{toomre64},  
\begin{equation}
Q=\dfrac{c_s\kappa}{\pi G \Sigma_0},
\label{mass_eq}
\end{equation} 
where $c_s$ is
the sound speed, $\kappa$ the epicyclic frequency,
and $\Sigma_0$ the background surface density. 
In a razor thin disk, linear axisymmetric
disturbances are unstable when $Q<1$, but
\emph{nonlinear} non-axisymmetric instability can occur for $Q\gtrsim
1$. If
the cooling is relatively inefficient, simulations show that
the disk falls into
a gravitoturbulent
state comprised of a disordered field of spiral density waves
that transports
significant angular momentum \citep{Gammie2001,rice14}. 
On the other hand, if the cooling is overly efficient the disc
fragments into dense clumps that may be
the precursors of gaseous giant planets or stars
\citep{cameron78,boss97}. 

Most numerical work is undertaken in a global setting, 
and indeed the characteristic lengthscales of the gravitoturbulent 
spiral waves is generally large-scale. But owing to inadequate 
resolution, global models struggle to describe dynamics on scales of
order the disk thickness $H$ or less, which are important for
turbulent mixing, dust settling, and (possibly) fragmentation. 
Local 3D shearing-box 
models are better suited for exploring these processes, though this
is a research direction that has been neglected, for the most part, in the
modelling of GI in disks. 
Recently, \citet{shi2014} presented 3D local box simulations that
showed that the gravitoturbulent state
failed to exhibit appreciable structure on
scales less than $H$ (see also \citet{hirose17}).  
However, their resolution (a maximum $8$ points per $H$) 
was probably too low to properly characterise any small-scale features.
In particular, it was unclear if their results had converged with
resolution. 
While the density waves are expected to exhibit
little vertical structure \citep[being essentially f-modes,][]{ogilvie98waves}, 
they are potentially vulnerable to parasitic instabilities
relying on a parametric resonance between small-scale inertial modes
\citep{fromang07c,bae16}. An open question is
whether this secondary instability also operates in gravitoturbulence,
and what influence it has on the properties of the spiral waves.

To address this issue, we performed 3D local shearing box
simulations of gravito-turbulent discs, with vertical stratification
and an ideal
gas law. Two different codes were used, PLUTO and RODEO, and
these were subject to a number of numerical checks, so as to 
confirm the trustworthiness of our results. 
Our main finding is that
vigorous and isotropic
small-scale turbulence develops slightly off the disk midplane 
and disturbs the spiral waves generated by GI. The
small-scale turbulence is obtained only in high resolution runs (at
least 20 cells per disc height scale), and thus has not been 
captured in previous under-resolved simulations: either SPH or
grid-based, 
local or global. 

We are persuaded that the small-scale activity is excited via a 3D parametric
instability that couples
pairs of inertial waves and a large-scale axisymmetric epicyclic
oscillation. Unstable inertial waves develop in the midplane but
break nonlinearly at a higher altitude, resulting in helical, incompressible
and non-axisymmetric motions around $z=0.5-1\,H$. We also consider
the merits of the Kelvin-Helmholtz instability, vertical splashing caused by
colliding density wakes, and the preferential
breaking of wakes in the disk atmosphere, but
find each in conflict with our numerical diagnostics. 
We discuss
the emergence of the large-scale epicyclic oscillations (upon which
the parametric instability feeds), and speculate on their importance to
both the subcritical transition to GI turbulence and its
continued sustenance.  

The structure of the paper is as follows: First, in
Section \ref{sec_model}, we introduce the basic equations of the problem
and discuss our numerical setup. In Section \ref{sec_prop}  we study the
dependence of the 3D gravito-turbulent state on the numerical
parameters,
in particular the grid resolution and box size. We
show evidence that turbulent quantities are converged with
horizontal box sizes $L_x=L_y$ if $L_x \gtrsim 40 H_0$.  In
Section \ref{sec_fourier}, we characterise the small-scale dynamics and the
large-scale axisymmetric oscillation in Fourier space. In
Section \ref{sec_parametric}, we analyse the nature of the
small-scale parasitic turbulence and show how it may result from a
parametric instability. Finally, in
Section \ref{sec_discussion}, we discuss the possible implications of
our results for disc physics, in particular for turbulent
mixing, dust dynamics, and the evolution of magnetic fields. 

\section{Numerical model}
\label{sec_model}
\subsection{{Governing equations}}

We assume that the gas orbiting around the central object is ideal, its pressure $P$ and
density $\rho$ related by $\gamma P=\rho c_s^2$, where $c_s$ is the
sound speed and $\gamma$ the ratio of specific heats. The pressure is
related to internal energy $U$ by $P=(\gamma-1)U$. We neglect 
molecular viscosity. We treat a local Cartesian model
of an accretion disk \citep[the shearing sheet;][]{goldreich65}. In this model, 
the axisymmetric differential rotation is approximated locally
by a linear shear flow and a
uniform rotation $\boldsymbol{\Omega}=\Omega \, \mathbf{e}_z$, with
$S=(3/2)\,\Omega$ for a Keplerian equilibrium.
We denote by $(x,y,z)$
the shearwise, streamwise and spanwise directions,
corresponding to radius, azimuth, and the vertical.
The shearing sheet approximation works best when the disk's angular
thickness is small \citep[see for e.g.][]{balbus99}, and thus
PP disks are only marginally covered by the model. 
In fact, for our
largest domains it is difficult
to justify the shearing box. 
AGN disks, being much thinner, are better described. 
We persist with the shearing sheet, however, as it provides a
well-defined platform 
and the superior
resolution needed to probe the smaller scales.

The evolution of density ${\rho}$, total velocity
{field} $\mathbf{v}$, and internal energy $U$ follows:
\begin{equation}
\dfrac{\partial \rho}{\partial t}+\nabla\cdot \left(\rho \mathbf{v}\right)=0,
\label{mass_eq}
\end{equation}
\begin{equation}
\frac{\partial{\mathbf{v}}}{\partial{t}}+\mathbf{v}\cdot\mathbf{\nabla
  v} +2\boldsymbol{\Omega}\times\mathbf{v} =-\nabla\Phi
  -\dfrac{\mathbf{\nabla}{P}}{\rho},
\label{ns_eq}
\end{equation}
\begin{equation}
\dfrac{\partial U}{\partial t}+\nabla\cdot (U\mathbf{v})
  =-P\nabla\cdot\mathbf{v}-\dfrac{U}{\tau_c}.
\label{energy_eq}
\end{equation}
 The total velocity field may be decomposed into
\begin{equation}
\mathbf{v}=-S x\,  \mathbf{e}_y+\mathbf{u},
\end{equation}
where $\mathbf{u}$ is the perturbed velocity field.  
The potential $\Phi$ is
 the sum of the tidal potential induced by the central object in the
 local frame, $\Phi_c=\frac{1}{2}\Omega^2 z^2-\frac{3}{2}\Omega^2\,x^2$, 
 and the gravitational potential induced by the disc itself, $\Phi_s$, 
 which obeys the Poisson equation: 
\begin{equation}
\nabla^2\Phi_s = 4\pi G\rho.
\label{poisson_eq}
\end{equation}
In the energy equation (\ref{energy_eq}), the cooling varies linearly
with $U$ with a typical timescale $\tau_c$ referred to as the `cooling
time'. This prescription is not very realistic but allows us to
control the rate of energy loss via a single parameter. We neglect
thermal conductivity. Finally, $\Omega^{-1}$ defines our
unit of time and $H_0$, the initial disc scale-height (defined below), 
our reference unit of
length.  

\subsection{Disc background equilibrium}

When $\tau_c\to \infty$, the governing equations admit an equilibrium
state characterised by $\mathbf{u}=0$, and vertical hydrostatic
balance. We consider homentropic equilibria, for which the vertical
profile
is polytropic $P=K\rho^\gamma$. Here
$K=c_{s_0}^2/(\gamma\rho_0^{\gamma-1})$, with $c_{s_0}$ and $\rho_0$
denoting the midplane sound speed and density, respectively. 
The equilibrium equations are
\begin{equation}
\label{eq_rho_eq}
K\left[\dfrac{1}{\rho}\dfrac{d\rho^\gamma}{dz}\right]+z\Omega^2+\dfrac{d\Phi_s}{dz}=0,
\end{equation}
\begin{equation}
\label{eq_phi_eq}
\dfrac{d^2\Phi_s}{dz^2}=4\pi G \rho,
\end{equation}
 which reduce to an inhomogeneous form of the Emden-Fowler
equation. 
Appendix \ref{appA} gives details on how to solve these equations
numerically. These solutions form part of our simulations' initial
condition.

\subsection{Numerical methods}

\subsubsection{Codes} 
\label{code}
Direct numerical simulations of the three-dimensional flow are
performed in the shearing box. The box has a finite domain of
size $(L_x,L_y,L_z)$, discretized on a mesh of $(N_X,N_Y,N_Z)$ grid
points. For most of the numerical runs, we use the Godunov-based PLUTO code
\citep{mignone2007}, which is well
adapted to highly compressible flow and shocks. This scheme uses a
conservative finite-volume method that solves the approximate Riemann
problem at each inter-cell boundary. It is known to successfully reproduce the behaviour of conserved quantities
like mass, momentum, and total energy. The
Riemann problem is handled by the HLLC solver which properly describes
contact discontinuities and has the advantage of being robust and positivity preserving. 

To allow longer time steps and eliminate numerical artifacts at the
boundaries, where the background shear flow is often very strong, we
use the orbital advection algorithm of PLUTO. It is based on splitting
the equation of motion into two parts, the first containing the linear
advection operator due to the background Keplerian shear and the
second the standard fluxes and source terms. Finally, PLUTO
conserves the total energy and so the heat equation is not
solved directly as in Eq.(\ref{energy_eq}). Hence the code captures the
irreversible heat produced by shocks due to numerical diffusion, 
consistent with the  Rankine-Hugoniot conditions. 

In Section \ref{code_comparison}, we compare our results with
another code, RODEO, that has previously simulated
self-gravitating disc in the 2D shearing box \citep{Pdkooper2012}. 
Similar to PLUTO, it is a Godunov-based method but one that relies on
the Roe solver \citep{roe81} to calculate interface fluxes. 
Second-order accuracy in space and time is achieved through a wave limiting
procedure as described, for example, in \cite{leveque02}. Most of the
results were obtained with the minmod wave limiter \citep{roe86},
although we briefly explore different limiters. One other significant
difference with PLUTO is that the RODEO simulations were performed in
\emph{shearing coordinates}, where the usual $y$ coordinate is
replaced by $y'=y+3/2\Omega t x$. In this way, the use of an orbital
advection algorithm is avoided, at the expense of a periodic remap
every $\tau_\mathrm{remap}$, which, by a clever choice of
$\tau_\mathrm{remap}$,  can be done by shifting an integer number
 of grid cells in $y$. This procedure therefore does not introduce any numerical diffusion. 

\subsubsection{Poisson solver}
\label{poisson_solver}

While in RODEO a pure Fourier Poisson solver was used \citep{koyama10},
a different approach was taken in the PLUTO simulations in order
to more robustly establish our numerical results. This is explained in
this subsection.

To compute the gravitational potential, we take advantage of the
shear-periodic boundary conditions. In a similar manner to
\citet{riols16a}, we first shift back the density in $y$ to the time
it was last periodic $(t = t_p)$. Then, for each plane of altitude
$z_k$, we perform the direct 2D Fourier transform of the density
$\rho$ and
obtain a vertical profile for each coefficient
$\hat{\rho}_{k_x,k_y} (z)$ of its Fourier decomposition, where $k_x$
and $k_y$ are radial and azimuthal wavenumbers. Using
Eq.~\ref{poisson_eq}, it is straightforward to show that in
Fourier space, the coefficients of the gravitational potential satisfy the Helmholtz equation:
\begin{equation}
\left[\dfrac{d^2}{dz^2} -k^2\right] \hat{\Phi}_{k_x,k_y} (z) = 4\pi G \hat{\rho}_{k_x,k_y} (z)
\label{eq_helmhotz}
\end{equation} 
with $k=k_x^2+k_y^2$, and 
$\hat{\Phi}_{k_x,k_y}$ the Fourier transform of the potential. 
This differential equation is solved in the
complex plane by means of a fourth-order 
finite-difference scheme and a direct
inversion method. The discretized system takes the form of a linear
problem $\mathsf{AX}=\mathsf{B}$ where $\mathsf{X}$ is a vector
representing the discretized potential's $z-$profile, 
$\mathsf{A}$ is a penta-diagonal matrix, and $\mathsf{B}$ is a column
vector containing the right hand side of Eq.~\ref{eq_helmhotz} and
extra coefficients setting the boundary conditions. We use a fast
algorithm involving $O(N_Z)$ flops to invert the matrix and obtain 
the discretized coefficents $\hat{\Phi}_{k_x,k_y} (z_k)$. 
For each altitude $z_k$, we finally compute the inverse Fourier
transform of the potential and shift it back to the initial sheared
frame. Note that gravitational forces are obtained by computing the
derivative of  the potential in each direction, 
using a 4th-order finite-difference method.

In total, the computational cost is of order $O(N\log(N_XN_Y))$ where
$N=N_XN_YN_Z$, and is hence less expensive than a full 3D Fourier
decomposition which would be $O(N\log(N))$. Moreover,
methods using vertical Fourier decomposition generally assume periodic
or vacuum boundary conditions for the potential. Hence, a correct
treatment of self-gravity in these methods requires the addition of two buffer
zones of size $L_z/2$ on either sides of the box, greatly increasing
the numerical workload
\citep{koyama10,shi2014}.  
In contrast, our setup can handle any boundary condition
without artificially augmenting the vertical domain. 
The stratified disc equilibria as well as the linear stability  
of these equilibria have been tested to ensure 
that our implementation is correct (see Appendices \ref{appA} and \ref{appB}). 

\subsubsection{Boundary conditions}

The shearing box framework implicitly assumes periodic boundary
conditions in $y$ and shear-periodic boundary condition in $x$. The
most delicate part is to assign suitable vertical boundary
conditions. We use the standard outflow conditions for velocity and
density fields but compute a hydrostatic balance in the ghost cells
for pressure, taking into account the large scale vertical component
of the self-gravity (averaged in $x$ and $y$). In this way we significantly reduce  
the excitation of waves near the boundary (see Fig.~\ref{fig_rhoprofile2} in Appendix \ref{appA}). 

For the gravitational potential, we impose 
\begin{equation}
\dfrac{d}{dz} \hat{\Phi}_{k_x,k_y} (\pm {L_z}/{2}) = \mp k\hat{\Phi}_{k_x,k_y} (\pm {L_z}/{2}).
\end{equation} 
This condition is an approximation of the Poisson equation in the limit of low density. Finally, we impose a density floor of $10^{-4}\, \Sigma/H_0$ which
prevents the timestep getting too small because of evacuated regions near
the vertical boundaries.

\subsection{Simulation setup}
\subsubsection{Box size and resolution}

The axisymmetric linear theory tells us that the fastest growing mode
possesses a radial lengthscale of order $H\,Q$ when $Q<1$.
Although our simulations focus on the
regime $Q\gtrsim 1$, we expect typical lengthscales
for spiral waves to be $\gtrsim H$. So in order to
obtain good statistical averages of the turbulent properties and ensure 
that the largest structures remain smaller than
the domain size, it is necessary to set $L_x,\, L_y \gg H$. 
As a compromise between this constraint
and numerical feasibility (for sufficient resolution),
we employ boxes of intermediate size $L_x=L_y=20 \, H_0$, though
we explore larger sizes in
Section \ref{resolution}.

 In most of the PLUTO simulations, we simulate a symmetric flow
 with respect to the midplane. 
 Thus, the vertical domain of the box only extends from 
the midplane to $3\,H_0$.
 We checked that 
anti-symmetric modes do not affect at all the properties of
 the turbulence and the results of the present paper. 

According to previous shearing box simulations
\citep{Gammie2001,shi2014}, a minimum resolution of 4-5 grid cells per $H_0$
in the horizontal directions is required
to correctly capture the onset of nonlinear instability. However,
to properly capture certain 
properties of the ensuing turbulence, as well as the fragmentation
criterion, more may be required. For example, \citet{Pdkooper2012} showed that in 2D, the
critical cooling time $\tau_c$ for fragmentation is still dependant on resolution when the
latter exceeds 40 points per $H_0$. In Section
\ref{resolution}, we compare different gravito-turbulent states
obtained at different resolutions, ranging from $3.2$ points per $H_0$
to almost $26$ points per $H_0$ in the horizontal directions.
In the vertical direction we normally used $64$ points in total over $3H_0$.

\subsubsection{Initial conditions and simulations parameters}
\label{ic}

In all our simulations we use a fixed heat
capacity ratio $\gamma=5/3$ and an initial midplane sound speed
$c_{s_0}=H_0/\Omega=1$. The initial conditions are similar to
those used in \citet{riols16a}, except that we must stipulate
vertical profiles for density and pressure. 

First, for an initial Toomre parameter $Q_0$ slightly larger than
1, we compute the vertical density and pressure profile associated with
a homentropic and self-gravitating disc equilibrium with no cooling (see
Eqs.~\ref{eq_rho_eq},  \ref{eq_phi_eq} and Appendix
\ref{appA}). 
Second, random non-axisymmetric density and velocity
perturbations are imposed upon this equilibrium. They possess a finite amplitude
decreasing exponentially with altitude. These initial fluctuations are
intensified by the instability and break down into a turbulent
flow after a short period of time $t\simeq 10 \,-\, 30 \,\Omega^{-1}$.

In order to prevent the disc from fragmenting in the early stages,
cooling is only introduced once the average turbulent quantities have
converged to a fixed value. This also provides a relatively 
`soft landing' onto the gravitoturbulent state, one that makes fewer demands
on the numerical scheme.  

If mass is lost through the
vertical boundary, it is replenished near the midplane. The
distribution of mass added to the disc at each time step exhibits a
Gaussian profile $\propto \text{exp}[-z^2/(2H_0^2)]$. The 
mass-injection rate varies in time so that $\Sigma_0=1$ remains
constant during the simulation. We checked that the amount of mass  
injected at each orbital period is negligible compare to the total
mass and 
does not affect the results. 

\subsection{Diagnostics}
\label{alpha}
\subsubsection{Averaged quantities}
To analyse the statistical behaviour of the turbulent flow, we define two different volume averages of a quantity $X$. The first one is the standard average:
\begin{equation}
\left<X \right>=\frac{1}{L_xL_yL_z}\int_V X\,\, dV.
\end{equation}
The second is the density-weighted average, defined by: 
\begin{equation}
\left<X \right>_w=\dfrac{\int_V \rho X\,\, dV}{\int_V \rho \,\, dV}.
\end{equation}
An important quantity that characterizes self-gravitating discs is the average 2D Toomre parameter defined by
\begin{equation}
Q_{2D}=\dfrac{\left\langle c_s\right\rangle_w \Omega}{\pi G \left\langle \Sigma\right\rangle },
\end{equation}
where  $\left<\Sigma \right>=L_z \left<\rho \right>$ is the average
surface density of the disc. To simplify notation, we 
denote $Q=Q_{2D}$.  

Another quantity that characterizes the turbulent dynamics is the
coefficient
 $\alpha$ which measures the angular momentum transport. This quantity is related to  
the average Reynolds stress $H_{xy}$ and gravitational stress $G_{xy}$.
\begin{align}
\label{def_alpha}
\alpha=\dfrac{2}{3\gamma \left\langle P\right\rangle}\left\langle
  H_{xy}+G_{xy} \right\rangle,
\end{align} 
where
\begin{align*}
H_{xy}=\rho u_xu_y \quad \text{and} 
\quad G_{xy}=\dfrac{1}{4\pi G}\dfrac{\partial \Phi}{\partial
  x}\dfrac{\partial \Phi}{\partial y}.
\end{align*}
The radial flux of angular momentum gives rise to the only source of energy in the system that can balance the cooling. This energy, initially in the form  of kinetic energy, is irremediably converted into heat by turbulent motions. 

Lastly, in order to study the energy budget of the flow, we introduce 
the average kinetic and internal energy denoted respectively by
\begin{align*}
 E_c=\frac{1}{2}\langle\rho \mathbf{u}^2\rangle,  \quad U=(\gamma-1)\langle P \rangle.
\end{align*}

\subsubsection{Fourier decomposition}
\label{decompo_fourier}
In Section \ref{sec_fourier}, we investigate the flow in
Fourier space. Any field $\mathbf{F}$ can be
decomposed in the following way:
\begin{equation}
\mathbf{F}=\! \sum_{\ell 
= -N_X/2}^{{N_X}/{2}}\,\,\,\, \sum_{m=-N_Y/2}^{{N_Y}/{2}}\!\!\hat{\mathbf{F}}_{\ell,m}(z,t)\exp\left[\text{i}(k_x(t)x+k_yy)\right]\;,
\end{equation}
with 
\begin{equation}
k_x(t)=\ell k_{x_0}+\frac{3}{2}\Omega \,m\, k_{y_0}t \quad \text{and} \quad k_y=m k_{y_0}
\end{equation}
The Lagrangian radial wavenumber $\ell k_{x_0}$ allows us to define
and label a given `shearing wave' $\ell$, if $k_y$ is known. This
quantity represents the wavenumber that a mode would have in a system
of coordinates following the shear. In the fixed coordinate system,
however, this wave has an Eulerian wavenumber $k_x(t)$ that increases
with time. The wave is referred to as `leading' when $k_y k_x(t)<0$
and `trailing'
 when  $k_y k_x(t)>0$. 

In practice, to compute the FFT transform of a given field, we first
have to express this field (in real space) in a system of coordinates
comoving with the shear, in a manner similar to our calculation 
 of the self-gravitating potential (see Section \ref{poisson_solver}). The change of variables is 
\begin{equation}
y \longrightarrow y' - \frac{3}{2}\Omega x (t-t_p),
\end{equation}
with $t_p$ the time corresponding to the nearest periodic point. This
process permits a strictly periodic field. We then compute the
standard FFT and obtain, for each altitude $z$ a spectral map in
$\ell$ and $m$. If we consider an interval of time $[nt_p,(n+1)t_p]$,
a point in this map represents the amplitude of a given  wave $\ell$
which has initially $k_x(nt_p)=\ell k_{x_0}$. However, this
representation is not valid for larger time, because the radial
wavenumber of the wave has increased during that time (especially for
large $k_y$) and one needs to remap the 2D Fourier spectrum onto a
grid in Eulerian wavenumbers ($k_x(t), k_y$). 

\section{Mean turbulent properties and numerical convergence}
\label{sec_prop}
In this section, we analyse the properties of 3D gravito-turbulence
and 
its dependence on resolution and box size for a fixed cooling time
$\tau_c=20\Omega^{-1}$. We then move on to explore the effects of
different initial conditions, numerical methods, and cooling times.

\subsection{Fiducial run: short and long term evolution}

\begin{figure*}
\centering
\includegraphics[width=0.94\textwidth]{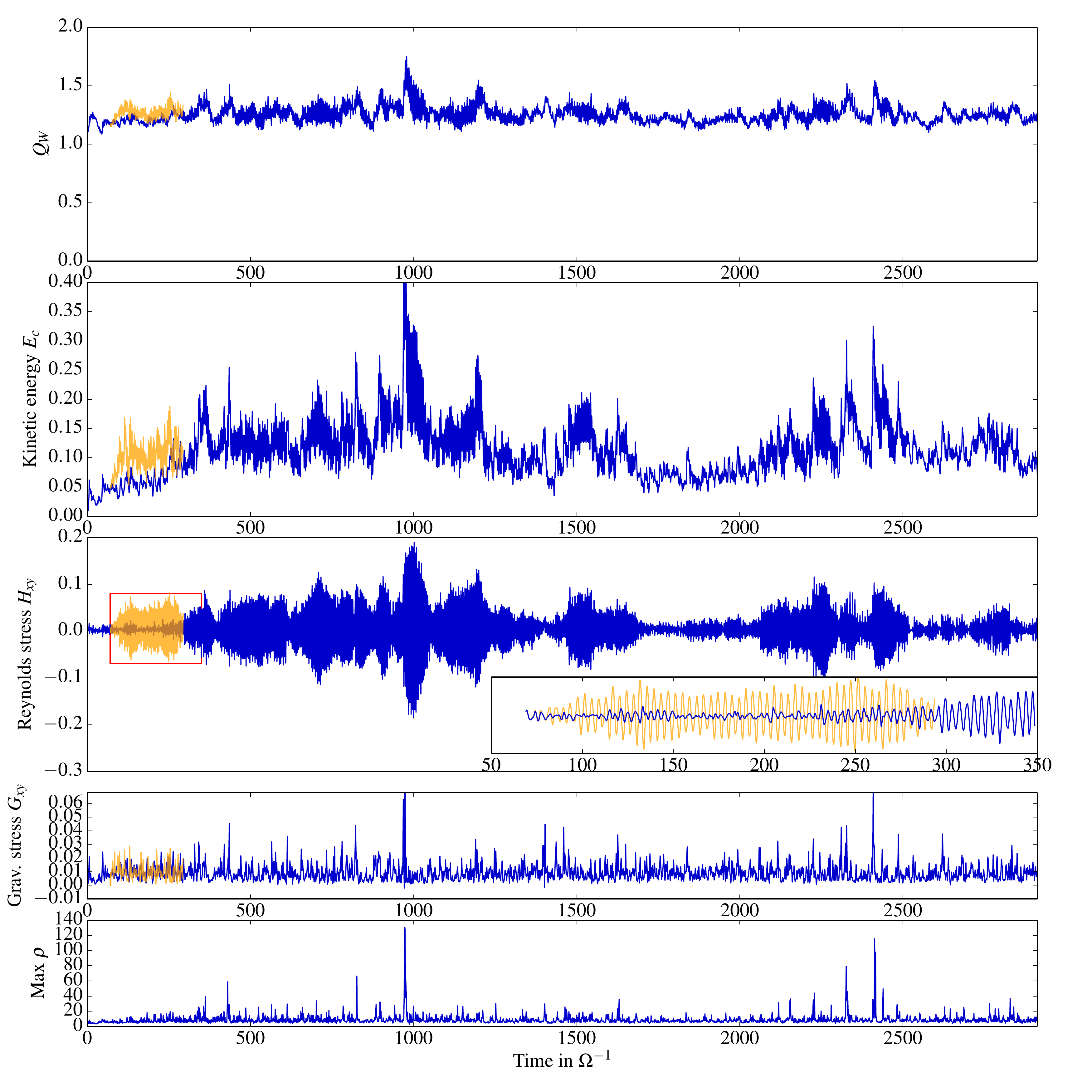}
 \caption{Time-evolution of various quantities, averaged over a box whose size is $L_x=20$, $L_y=20$ and $L_z=3 \,H_0$. From top to bottom, density-weighted average Toomre parameter $Q$, box average kinetic energy, Reynolds stress, gravitational stress, and density maximum. The resolution is $128\times128\times 64$ (run PL20-128).}
\label{fig_average}
 \end{figure*}

We first explore the gravito-turbulent flow on long time
scales. 
To that end we ran a simulation, labelled
PL20-128, with horizontal resolution similar to
that of \citet{shi2014} ($\approx 6.5$ cells per $H_0$) in a box of
size $L_x=L_y=20$ for almost 3000 $\Omega^{-1}$. This simulation will be considered as our reference
test run. Some average turbulent properties are summarized in Table
\ref{table1} and their evolution in time shown in Figure
\ref{fig_average} (blue/dark curves). 

A quasi-steady turbulent state
as well as a thermodynamic equilibrium is obtained within a
few tens of orbits, characterised by $Q\simeq 1.24$,
very similar
 to the value calculated by
 \citet{shi2014}. We verified that the average transport efficiency
 $\alpha$, defined in Eq.~(\ref{def_alpha}) matches the
 prediction of  \citet{Gammie2001},
\begin{equation}
\label{eq_gammie}
\alpha=\frac{1}{q^2\Omega \tau_c \gamma (\gamma-1)},
\end{equation}
which is based on total energy conservation.
The average kinetic energy associated with the fluctuations remains
smaller than the average internal energy, as in classical 2D razor
thin disc simulations. The gravitational stress is always
positive and contributes to most of the angular momentum transport,
while the Reynolds stress $H_{xy}$ is subdominant, in agreement with 
\citet{shi2014}.  We also checked that these results do not depend on
initial conditions. For that purpose, we ran a simulation PL20-128-b, 
initialized with different noise realization (i.e different Fourier amplitudes). This simulation exhibits similar $Q$ and mean average properties as PL20-128. 

While all mean quantities fluctuate on a timescale of several orbits, 
Figure \ref{fig_average} shows that
$H_{xy}$ and kinetic energy $E_c$ undergo significant variations on
a much longer timescale, of order $\sim 1000 \,\Omega^{-1}$.
They
exhibit bursts of activity, for example 
between $t=500$ and $t=1700$ $\Omega^{-1}$,
followed by quiescent phases in which the kinetic energy can be three times
smaller (such as between $t=1700$ and $t=2200$ $\Omega^{-1}$). The last
panel of Fig.~\ref{fig_average}  shows that bursts of activity are
generally associated with the formation of transient clumps where the
local density can exceed 100 times the background density. These 
transient clumps do not collapse during the simulation but could 
be important in the process of stochastic fragmentation
\citep{Pdkooper2012}.  
We conclude that this long-time behaviour is not driven by the
correlation
length of the turbulence reaching the box size, and is hence physical
and not an artefact of the shearing box model.
However, it does make it extremely difficult to obtain meaningful
averages for $E_c$ and $H_{xy}$ because simulations must be run for
extremely long.

 Although the long-term evolution of The Reynolds stress $H_{xy}$ is stochastic and bursty, on
 short time it oscillates quasi-periodically between negative and positive values with a  frequency close to the orbital frequency $\Omega$ (on average
 it has a positive contribution). These fast and regular oscillations
  are also visible in the simulations of \citet{shi2014}. 
 In Sections \ref{sec_fourier} and
 \ref{sec_parametric}, we inspect these
 oscillations in more detail 
and show that they control several aspects of the dynamics.

\begin{figure}
\centering
\includegraphics[width=\columnwidth]{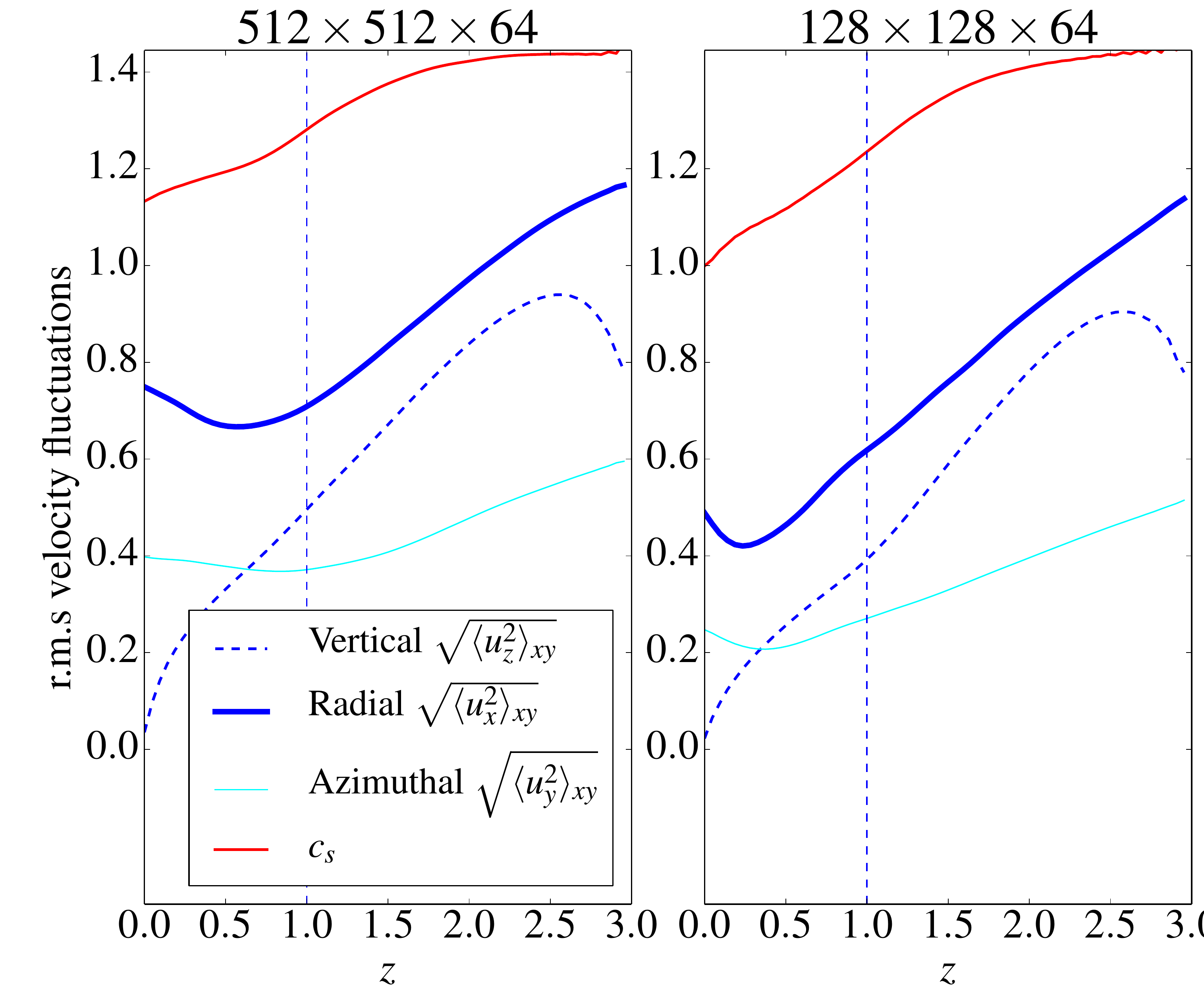}
 \caption{Vertical profiles of turbulent r.m.s velocities, time
   averaged over 
 $40\, \Omega^{-1}$ for the run PL512 at high resolution (left) and 
 for the run PL128 at low resolution (right)}
\label{fig_rms}
 \end{figure}

 Finally, we analysed the time-averaged r.m.s fluctuations as a
 function of altitude $z$. Fig \ref{fig_rms} shows that for our test
 reference simulation (right panel), the radial turbulent velocity
  is always larger than the other components, and in fact
 is roughly twice the azimuthal component, a signature of
 large-scale density waves. As $z$
 increases, the ratio between vertical and radial r.m.s fluctuations
 increases, and approaches 1 as $z\sim H_0$. This trend also appears
 in
  \citet{shi2014}. The non-negligible value of $v_z$ in the atmosphere
  is possibly due to the `vertical breathing' motion of spiral waves,
  characteristic of f-modes in polytropic or self-gravitating
  isothermal gas \citep[AppendixB]{kory95,ogilvie98waves}. The vertical velocity may also be
  enhanced by `vertical splashing' as spiral density waves collide.
  We note that the average sound speed
 (or temperature) increases slowly with $z$ and is always greater than each
 velocity component taken individually, and thus the mean motions are
 slightly subsonic. We discuss in the next
 subsection how these different results depend on grid resolution and box size. 

\begin{figure}[$L_x=L_y=20 \,H_0$]
\centering
\includegraphics[width=\columnwidth]{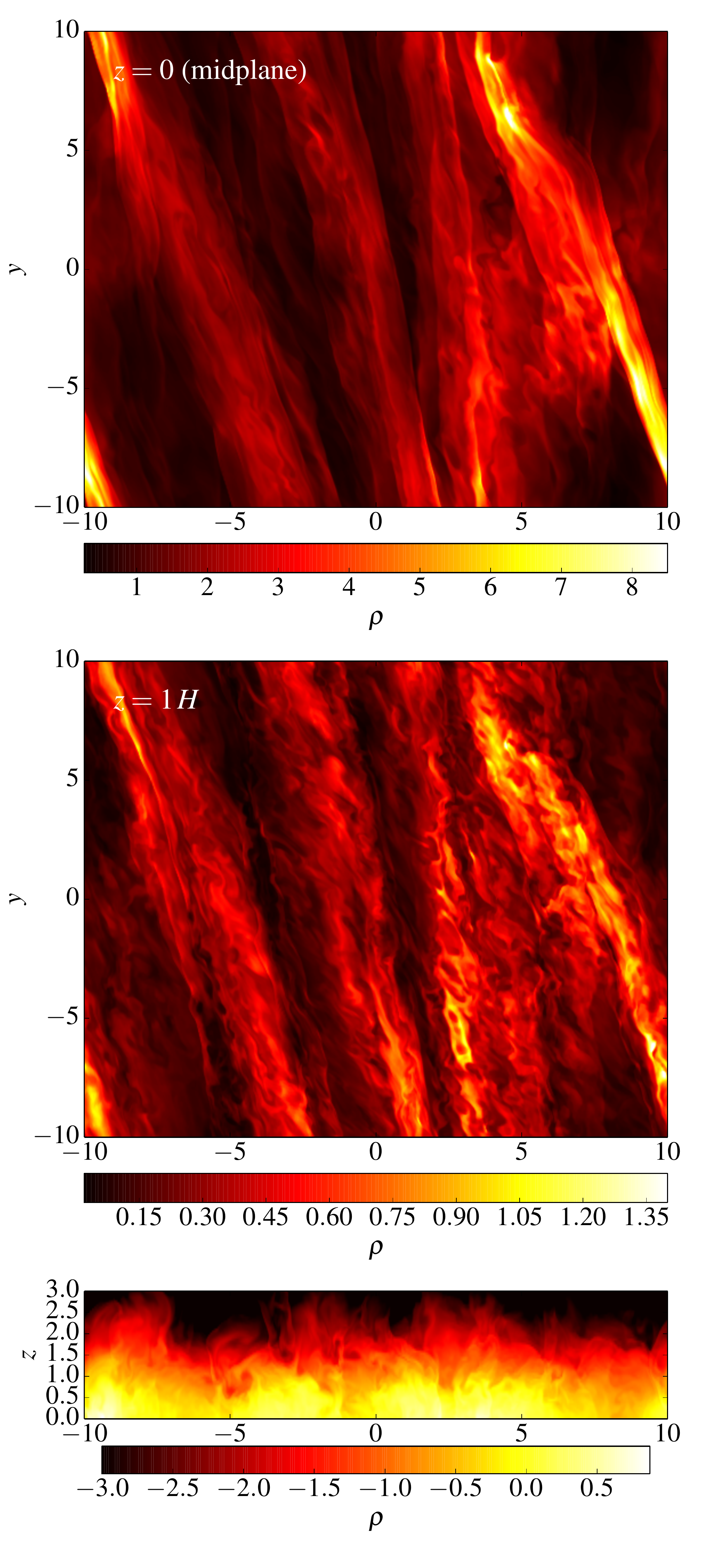}
 \caption{Density snapshots of the 3D gravito-turbulent state obtained
   in a box of $L_x=L_y=20 H_0$ and resolution of $\approx 25$ cells per $H_0$
   with $\tau_c=20\, \Omega^{-1}$. The top panel is a view of the disc
   midplane $z=0$, the central panel is for $z=H_0$, and the bottom
   panel is a vertical cut, with $\rho$ represented in a logarithmic scale.}
\label{fig_rhomap}
 \end{figure}
 \begin{figure}[$L_x=L_y=40 \,H_0$]
\centering
\includegraphics[width=\columnwidth]{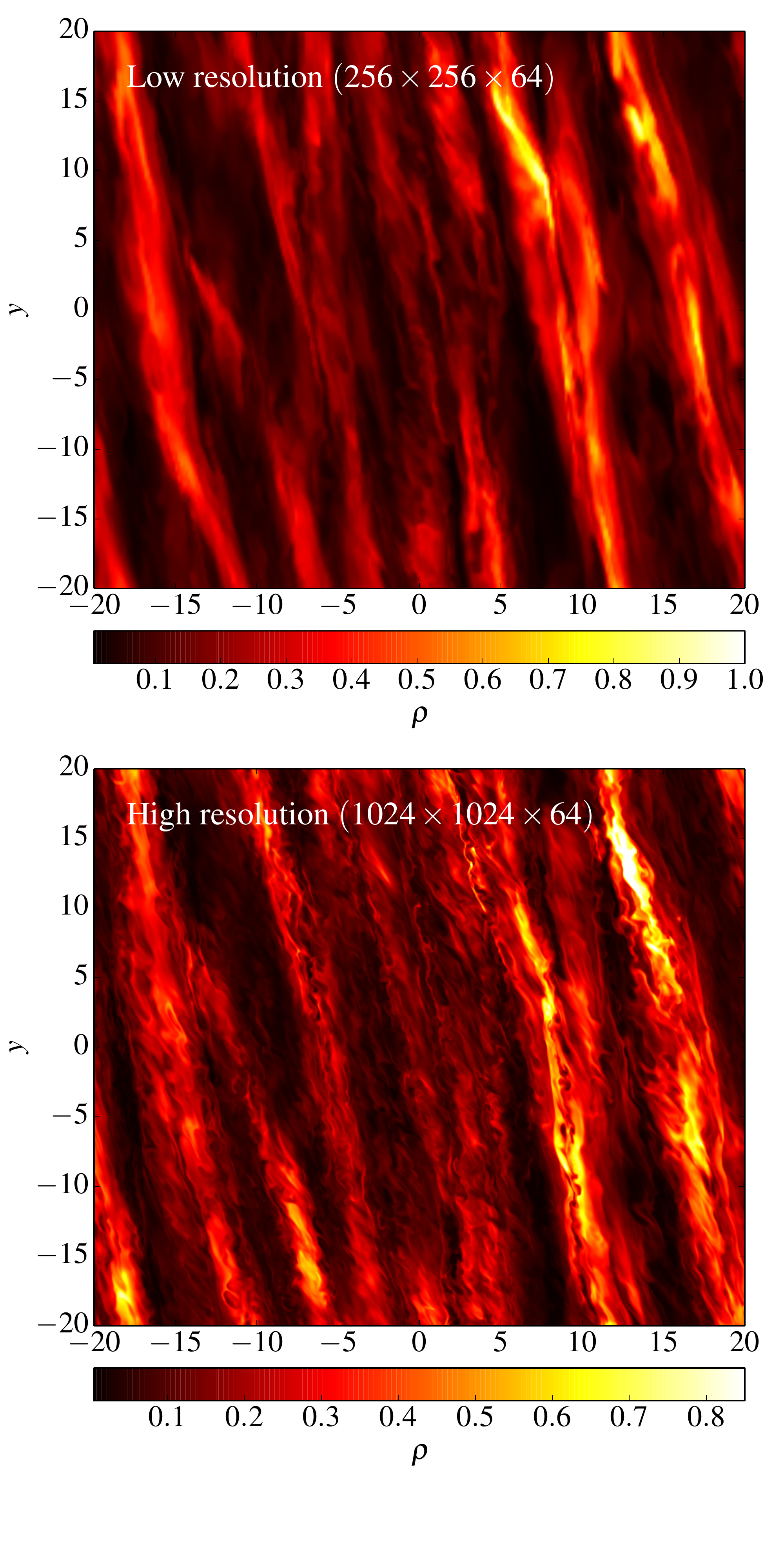}
 \caption{Density snapshots of the 3D gravito-turbulent state obtained
   in a box of $L_x=L_y=40 \,H_0$ at altitude $z=H_0$. The top panel
   comes from a low resolution run of $\approx 6.5$ cells per $H_0$
   (PL40-256),
    whereas the bottom panel comes from a high resolution run, $\approx 25$ cells per $H_0$ (PL40-1024). }
\label{fig_rhomap2}
 \end{figure}

\subsection{Dependence on resolution}
\label{resolution}

Next we study how the average turbulent properties depend on
grid resolution. For a fixed box size $L_x=L_y=20 \,H_0$, we compare
four different setups, detailed in Table \ref{table1}, with  64, 128,
256, 512 points in the horizontal directions (ranging from 3.2 points
to $25.6$ points per $H_0$.).

Table \ref{table1} shows that
some quantities like the gravitational stress and Toomre parameter
$Q$ do not vary significantly as resolution is increased, provided
that $N_X\geq 128$. In contrast, the kinetic energy and, in particular, the Reynolds
stress differ from one resolution to another, because
the higher resolutions simulations have not been run
sufficiently long. Recall that in the previous
subsection these quantities were shown to vary on periods of thousands
$\Omega^{-1}$, but such run times are inaccessible at high resolution
(the highest resolved simulations have been run for  $\sim 200 \,
\Omega^{-1}$). The values listed for the kinetic energy and Reynolds stress are
probably statistically insignificant. 

\begin{center}
\begin{table*}    
\centering 
\begin{tabular}{c c c c c c c c c c}          
\hline                        
Run & Resolution & Time (in $\Omega^{-1}$) & $\tau_c$ & $Q$ & $E_c$ & $U$ & $H_{xy}$ & $G_{xy}$ & $\alpha$\\    
\hline  
	PL20-64 	& $64\times 64\times 32$ & 500 & 20 & 1.14 & 0.047 & 0.332 & 0.00198 & 0.00751 & 0.0176\\
	\textcolor{red}{PL20-128}	& $128\times 128\times 64$ & 3000 & 20 & 1.24 & 0.105 & 0.396 & 0.00411 & 0.00915 & 0.02005\\                
    PL20-256 &  $256\times 256\times 64$ & 100  & 20 & 1.25 & 0.115 & 0.427 & 0.005 & 0.0109 & 0.022\\      
    PL20-512 & $512\times 512\times 64$ &  200 & 20  & 1.26 & 0.109 & 0.412 & 0.00594 & 0.00963 &  0.0226\\
    \hline
    PL20-128-b & $128\times 128\times 64$ & 400 & 20 & 1.22 & 0.0762 & 0.377 & 0.00376 & 0.00977 & 0.0212\\              
    PL20-256-RK3 &  $256\times 256\times 64$ &  200 & 20 & 1.23 & 0.0741 & 0.384 & 0.00509 & 0.00955 & 0.0229\\      
     PL20-512-$\beta10$ & $512\times 512\times 64$ &  200 & 10  & 1.27 & 0.126 & 0.382 & 0.0091 & 0.0157 &  0.039\\
    \hline 
    RO20-128	& $128\times 128\times 128$ & 400 & 20 & 1.18 & 0.0777 & 0.472 & 0.00419 & 0.0112 & 0.0198\\                             
    RO20-264 &  $264\times 264\times 128$  & 200 & 20 & 1.20 & 0.08 & 0.489 & 0.00486 & 0.0124 & 0.0212\\
    RO20-264-FL2 &  $264\times 264\times 128$  & 200 & 20 & 1.19 & 0.081 & 0.475 & 0.00539 & 0.0118 & 0.0217\\
    RO20-512 &  $512\times 512\times 128$  & 90 & 20 & 1.18 & 0.0754 & 0.464 & 0.00488 & 0.0116 & 0.0213\\                              
\end{tabular}  
\vspace{0.5cm}
\caption{Parameters and properties of runs in a box of $L_x=L_y=20 H_0$. The third columm indicates the time over which quantities have been averaged (not including the transient phase). $Q$ is the average Toomre parameter, $E_c$ and $U$ are the box and time-averaged kinetic and internal energy, $H_{xy}$, $G_{xy}$ and $\alpha$ are respectively the averaged Reynolds, gravitational stress and transport efficiency.}  
\vspace{0.5cm}
\label{table1}
\end{table*}  
\end{center}   
\begin{center}
\begin{table*}    
\centering 
\begin{tabular}{c c c c c c c c c c}          
\hline                        
Run & Resolution & Time (in $\Omega^{-1}$) & $\tau_c$ & $Q$ & $E_c$ & $U$ & $H_{xy}$ & $G_{xy}$ & $\alpha$\\    
\hline  
	\textcolor{red}{PL40-256} 	& $256\times 256\times 64$ & 100 & 20 & 1.38 & 0.0753 & 0.4 & 0.00249 & 0.0114 & 0.02115\\
		
	PL40-512	& $512\times 512\times 64$ & 70 & 20 & 1.40 & 0.132 & 0.524 & 0.0034 & 0.0186 & 0.025 \\                             
    PL40-1024 &  $1024\times 1024\times 64$ &  50 & 20 & 1.36 & 0.076 & 0.4 & 0.0028 & 0.012 & 0.023   \\
     \hline 
    RO40-256	& $256\times 256\times 128$ & 900 & 20 & 1.36 & 0.117 & 0.625 & 0.00281 & 0.0213 & 0.0233\\
\end{tabular}  
\caption{Simulations in larger boxes, $L_x=L_y=40 H_0$.} 
\vspace{0.5cm}
\label{table2}
\end{table*}  
\end{center}  
\begin{center}
\begin{table*}    
\centering 
\begin{tabular}{c c c c c c c c c c}          
\hline                        
Run & Resolution & Time (in $\Omega^{-1}$) &$\tau_c$ &  $Q$ & $E_c$ & $U$ & $H_{xy}$ & $G_{xy}$ & $\alpha$\\    
\hline  
	\textcolor{red}{PL80-512} 	& $512\times 512\times 64$ & 100 & 20 & 1.41 & 0.131 & 0.523 & 0.0032 & 0.0182 & 0.0244\\  
	\hline
	RO80-512 	& $512\times 512\times 128$ & 200 & 20 & 1.39 & 0.149 & 0.664 & 1e-6 & 0.0272 & 0.0246\\        
\end{tabular}   
\vspace{0.5cm}
\caption{Simulations in large boxes, $L_x=L_y=80 H_0$.}
\label{table3}
\end{table*}  
\end{center}  

Figure \ref{fig_rms} shows the vertical profile of r.m.s velocity
fluctuations for PL20-128 and PL20-512 (averaged over
$40\Omega^{-1}$). Increasing resolution does not seem to affect the
balance between each spatial component, although the radial and
azimuthal fluctuations are 1.5 stronger with a resolution of
$512$ points in $L_x$ and $L_y$. 
We note that vertical motions remain important at
$z \geq H_0$ whatever the resolution 
used (even for the lowest one, $N_X=64$). 

 Finally, we visually
examined the density snapshots at different altitudes and
resolutions. Fig.~\ref{fig_rhomap} shows the density structures
in the midplane and one scale-height above. At $z=0$, the turbulence
is comprised of large scale and distinct spiral waves, of
radial lengthscale $\approx 4-5 H_0$, which break
non-linearly. However, at larger $z\simeq H_0$ and in higher
resolution runs, the coherence of the
spiral waves is disturbed by a form of small-scale
turbulence. This does not appear in our low resolution runs
PL20-128, PL40-256 or PL80-512, nor in previous
simulations of 3D gravito-turbulence. Figure \ref{fig_rhomap2} shows, in
particular, a comparison between different resolutions at $z=H_0$ (and
for a box of sixe $L_x=40 \,H_0$). Visually these small scale motions
take the form of wispy deformations of the spiral wave fronts. These small-scale
fluctuations are relatively strong (of order the host density wave),
highly non-axisymmetric, and mainly
located between $z=0.5\, H_0$ and $z=H_0$. The density at that
location is still relatively high, varying between 0.1 and 1. These small scale
features remain visible throughout the simulation and are observed
independently of the code, boundary conditions (see
section \ref{par_code}) and box size (see Fig.~\ref{fig_rhomap} and Fig.~\ref{fig_rhomap2}
 for comparison). 

In conclusion, there is evidence that we achieve convergence 
for certain mean gravito-turbulent
quantities ($Q$ and $G_{xy}$ notably)
with resolution, provided that we use more than
$5$ grid cells per $H_0$ in $x$ and $y$. We cannot decide on the
convergence of the kinetic energy $E_c$ and
Reynolds stress $H_{xy}$, because they exhibit strong fluctuations on
times of several hundreds of orbits, of order or longer than our highest
resolution simulations. 
Finally, at resolution larger
than $20$ cells per $H_0$, we discover a form of small-scale turbulence
attacking the large-scale spiral waves. The behaviour of this new
dynamical feature with resolution is probably still unconverged.
And though its influence on the mean turbulent quantities is probably
mild (possibly appearing in the vertical profiles of $u_x$ and
$u_y$), 
it is readily identified visually in snapshots of the density
field, and in later section we show how it can be quantified 
through various diagnostics.

\subsection{Dependence on box size}

Another important test, especially for local models,
is the box-size dependence of the turbulent
properties.
 Table \ref{table1}, \ref{table2} and \ref{table3} show,
respectively, a set of simulations computed for $L_x=20$, $40$ and $80$
$H_0$. To aid comparison, fiducial runs at different box sizes but the
same resolution per $H_0$ are highlighted in red.

First, doubling the horizontal size from 20 to 40 $H_0$
while keeping the resolution per scale height fixed, modifies
the average $Q$ only moderately. It is almost 15\% larger in PL40-256
compared with PL20-128. This result seems to be robust since it is
obtained at different resolutions and with different codes (see
Section \ref{code_comparison}). The internal
energy also increases. But more
strikingly, the ratio between the gravitational and the Reynolds
stress is multiplied by a factor 2 or even 3. 
Unlike the variations of $H_{xy}$ with resolution, this result is
statistically robust, since quantities have been recorded for at least
900 $\Omega^{-1}$ (run RO40-256) and  
the same trend is obtained for all simulations with $L_x=40\, H_0$. Oscillations of $H_{xy}$
at a frequency of $\Omega$ are still clearly discernible, although
their amplitude are weaker and rarely reach negative values. 

Finally,
for a box even larger, of size $L_x=L_y=80\, H_0$,  we show that
turbulence properties are comparable to those obtained for $L_x=L_y=40\,
H_0$. In particular it appears $Q$ has settled down to a
value $\simeq 1.35$-$1.4$. Discrepancies in kinetic energy and
Reynolds stress are probably due to insufficient statistics. In
conclusion,
we have some evidence that
there exists some critical $L_x=L_y$ between 20 and 40 $H_0$ above which
simulations are converged with respect to the box size. However,
further long-term simulations are needed to nail this down.

\subsection{Dependence on cooling time}

We ran a small number of simulations with different cooling times,
not enough to ascertain a critical rate at which fragmentation occurs,
but to fix some basic ideas.
A high resolution simulation, PL20-512-$\beta$10,
using $\tau_c=10\,\Omega^{-1}$ and run for $200\,\Omega^{-1}$
shows that $Q$ does not vary significantly compared to when
$\tau_c=20\,\Omega^{-1}$ (with same resolution). However the average
stress is multiplied by a factor 2, as expected from
Eq.~\ref{eq_gammie}. Note that, despite the shorter cooling time,
this simulation failed to exhibit collapsing or even transient
fragments.
We did check however that when $\tau_c=3\,\Omega^{-1}$
the system fragments in less than 20 orbits which supports
the results of \citet{shi2014}.

\subsection{Code comparison and dependence on numerical details}

\label{par_code}
To verify that our results are robust and that the simulated
gravito-turbulence does
not depend strongly on our numerical implementation, we compare our
results with a different code, RODEO.
It employs a different Riemann solver, Poisson
solver, and boundary conditions (see Sections
\ref{code} and \ref{poisson_solver} ). 

Tables \ref{table1}, \ref{table2} and
\ref{table3} show relatively good agreement between the two codes
(within the constraints of run-time), 
in
particular for the value of mean $Q$. There are nevertheless some
differences: the internal energy is systematically larger in RODEO than
in PLUTO. Moreover, in the $L_x=80 \, H_0$ runs, $G_{xy}$ is also
larger. 
These difference remain relatively small and  
could be due to insufficiently long integrations.  

We also checked that simulations 
run with RODEO depend only mildly on the flux
limiter used (see comparison on Table \ref{table1} between RO20-264
and RO20-264-FL2). Finally, we show that results are independent of
the order of the Runge-Kutta method in  
the time integrations (see runs PL20-256 and PL20-256-RK3).

\label{code_comparison}

\section{Small-scale dynamics and axisymmetric oscillations}
\label{sec_fourier}

We showed previously that gravito-turbulent states are
characterized by regular oscillations of the Reynolds stress and, in
the best resolved simulations, vigorous
small-scale dynamics around $z\simeq H_0$. 
In this section, we better understand
these structures by extracting their signatures in Fourier space.  

\subsection{Turbulent power spectra and small-scale structures}
\label{smallscaledyn}

We
denote by $\widehat{\mathbf{u}}^w(k_x,k_y,z,t)$ the horizontal 2D
Fourier transform
of the weighted turbulent velocity field $\rho^{1/3}\mathbf{u}$ 
(see Section \ref{decompo_fourier} for definitions).  
The density-weighted, time-averaged,
2D kinetic energy power spectrum may then be defined through
\begin{equation}
E(k_x,k_y,z)=\frac{\Sigma^{1/3} H_0^{-1/3}}{2 T}
 \int_{t_0}^{t_0+T}\left\vert
   {\widehat{\mathbf{u}}^w(k_x,k_y,z,t)}\right\vert^2 \,dt,
\end{equation}
where the time-average begins at $t=t_0$ and is carried out over an
interval of duration $T$.

 The weighted
spectrum, with the one-third scaling in density, is often used
in the study of supersonic and compressible turbulence
\citep{Lighthill55,kristuk07}. Using simple scaling arguments, one
can show that $E(k)\sim k^{-5/3}$ and hence follows the
incompressible Kolmogorov power law. Note, however, that this
scaling fails for  
highly compressible forcing  and cannot be considered 
universal \citep{federath13}.  

We define next the time and $k_x$-averaged 1D spectrum 
\begin{equation}
E_{\text{1D}}(k_y,z)=\dfrac{1}{k_{x_0}}\int E (k_x,k_y,z)\,d k_x,
\end{equation}
and the time-averaged isotropic 1D spectrum 
\begin{equation}
E_{\text{iso}}(k,z)=\int \int E (k_x,k_y,z)\, \delta(| \mathbf{k}'|-k) \,d k_x dk_y,
\end{equation}
where $|\mathbf{k}'|=\sqrt{(k_x')^2+(k_y')^2}$ is the radius in
horizontal wavenumber
space. The $k_x$-averaged spectrum is particularly useful in
distinguishing the GI wakes (which possess a low $k_y$) from small-scale
turbulence (for which $k_y$ is bigger). Note that both features may possess 
comparable $k_x$.

\begin{figure}
\centering
\includegraphics[width=\columnwidth]{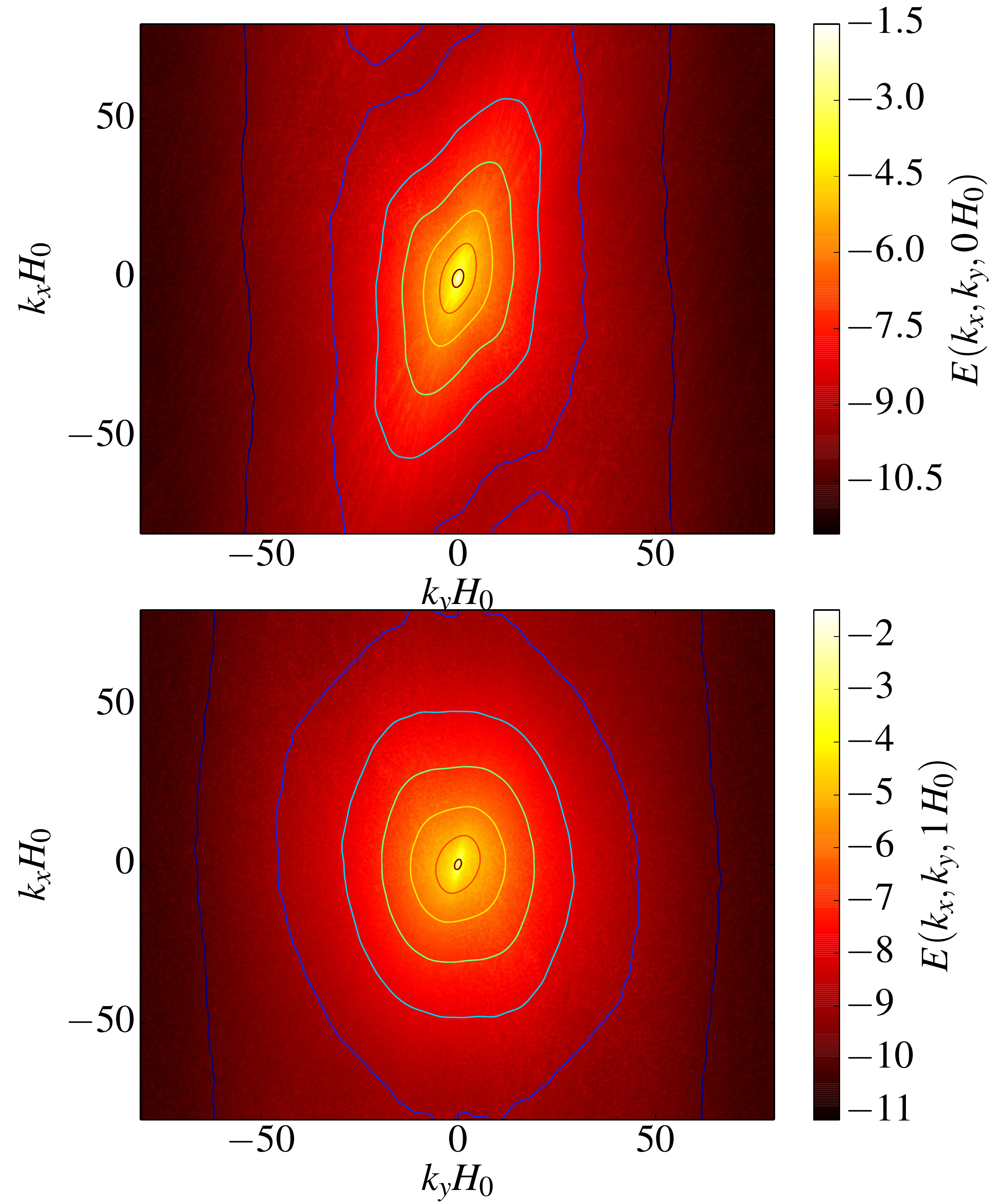}
 \caption{2D horizontal kinetic energy power spectrum $E(k_x,k_y,z)$
  as a function of Eulerian wavenumbers in simulation PL20-512.
 The spectrum is averaged over  $40\, \Omega^{-1}$. The top panel is for $z=0$ and the bottom panel is for $z=H_0$.}
\label{fig_spec1}
 \end{figure}
 \begin{figure}
\centering
\includegraphics[width=\columnwidth]{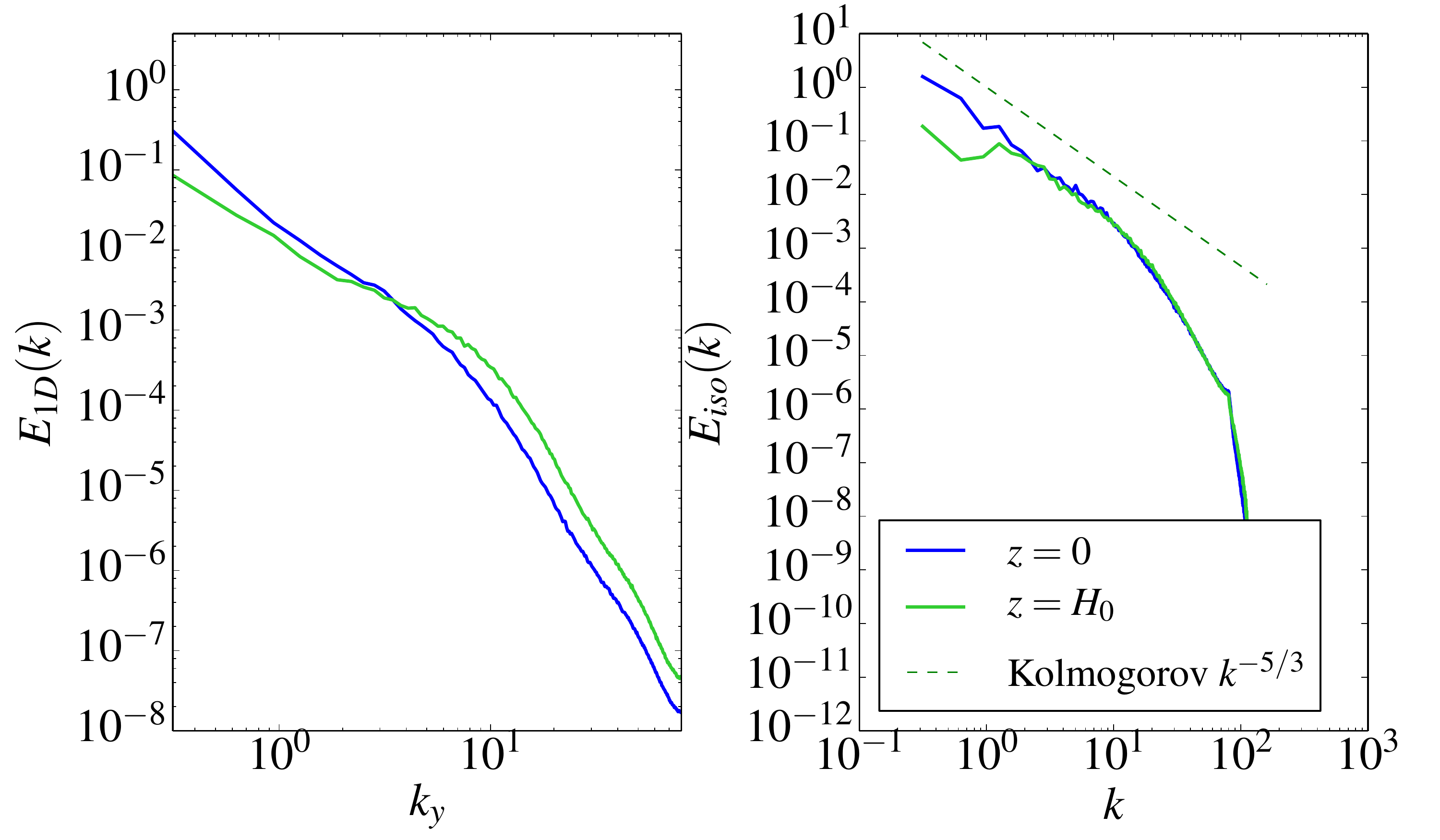}
 \caption{Left: 1D kinetic energy spectrum $E_{\text{1D}}$. Right: 1D
   isotropic spectrum $E_{\text{iso}}$. 
   Two different altitudes have been considered: blue curves are for $z=0$ and green curve are for $z=H_0$. Data are computed from the simulation PL20-512 and averaged in time over $40\Omega^{-1}$.  }
\label{fig_spec2}
 \end{figure}

Figure \ref{fig_spec1} shows the time-averaged 2D power spectrum of
gravito-turbulence $E(k_x,k_y,z)$ as a function of the
horizontal Eulerian wavenumbers at two different altitudes $z=0$ and
$z=H_0$. Both spectra are computed from high resolution simulation
data (PL20-512) and averaged over $40\, \Omega^{-1}$. For $z=0$, the
energy is contained in an inclined elliptical band along the $k_y=0$
axis. Turbulent structures are weakly non-axisymmetric and elongated
along the azimuthal direction with a small pitch angle. In contrast, at
$z=H_0$ the signal appears far more isotropic; energy is clearly
spread onto small scale non-axisymmetric modes. This broadening of the
spectrum is undoubtedly related to the small-scale disturbances afflicting
the spiral waves described earlier in Section
\ref{resolution}.  

To better distinguish the small scale motions,  
we plotted $E_{\text{1D}}(k_y,z)$, the $k_x$-averaged 1D
spectrum, 
in the left panel of Fig.~\ref{fig_spec2} evaluated at $z=0$ and
$z=H_0$. Note that the ratio of large-scale energy ($k_y\leq
2\pi/L_y$) to small scale energy ($k_y=50$ for instance) strongly varies
with altitude $z$. At $z=H_0$ this ratio is 15 times
smaller than at $z=0$, indicating a significant distribution of energy to
smaller scales higher up in the disk. The spectrum in
$k_y$ appears also less steep for azimuthal scales larger than $0.5
H_0$, which suggests that the energy of non-axisymmetric modes is injected
and cascades differently at those two altitudes.  Lower resolution runs
do not exhibit nearly the same amount of power on the resolved
small-scales at $z=H_0$.

Lastly, out of interest we plot
the isotropic spectrum $E_{\text{iso}}(k,z)$, even though it is
difficult
to distinguish the two features in it because both share similar $k_x$.
Note, however, that
the slope  of $E(k)$ follows a Kolmogorov law for $k<10 H_0^{-1}$ 
but then appears much steeper (well before approaching the grid scale).

In conclusion, the small-scale parasitic turbulence
attacking at $z=H_0$ is
quantifiable and possesses a clear signature in the Fourier
decomposition of the flow. This supports the conclusions of the
previous
section obtained by visual inspection of the density maps.
 In light of Figs \ref{fig_rhomap},~\ref{fig_rhomap2}  and
the present results, we assume that the mechanism producing those
small-scale features could be more efficient than a simple inertial
turbulent cascade and could be potentially due to a secondary
instability of the 
large scale structure. 

\subsection{Large scale modes: spiral waves and axisymmetric
  epicyclic oscillations}

The turbulent power spectra shown in Fig.~\ref{fig_spec2} reveals that
kinetic energy is not concentrated at wavelengths $\approx 4-5H_0$,
the typical size of GI density wakes. Apart from the short scales, energy
keeps increasing to very large scales $\approx L_x$. In this
section, we investigate in more detail the nature and 
 origin of these large scale features. 

\begin{figure}
\centering
\includegraphics[width=\columnwidth]{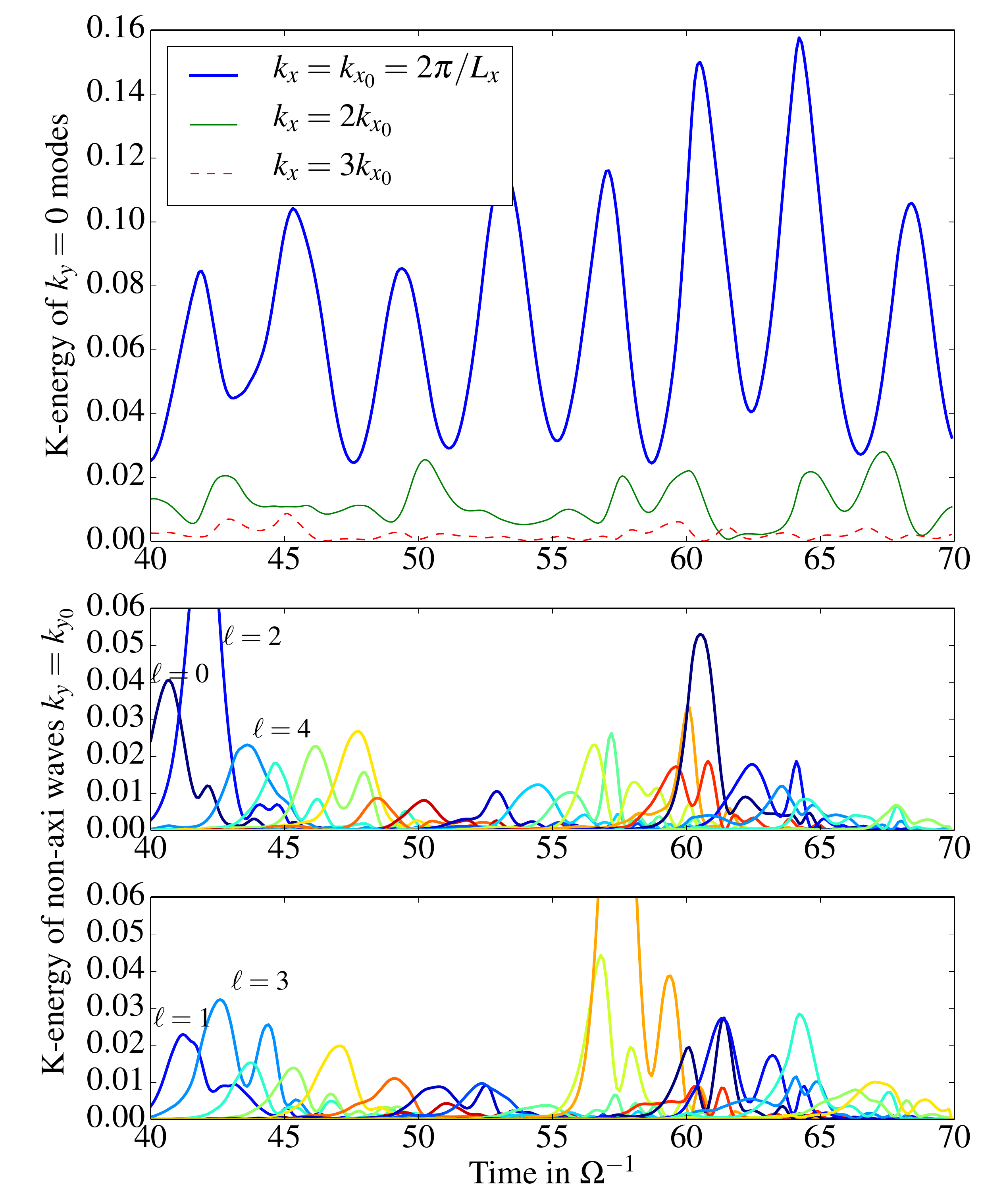}
 \caption{Top: Time-evolution of the kinetic energy $E (k_{x_0},0,0)$,
   $E (2k_{x_0},0,0)$ and $E (3k_{x_0},0,0)$  associated with the
   first few
   axisymmetric modes. Bottom:  kinetic energy of several
   non-axisymmetric shearing waves, with $k_y=2 \pi/L_y$, labelled by
   their Lagrangian wavenumber $\ell$ (to improve readability, we
   separate the even $\ell$ in the centre panel from the 
  odd $\ell$ in the bottom panel). 
   All the spectral quantities have been computed in the midplane $z=0$.}
\label{fig_energy}
 \end{figure}

Figure \ref{fig_energy} shows the time-evolution of kinetic energy
associated with large scale $m=0$ axisymmetric modes (top panel) and $m=1$
non-axisymmetric shearing waves (bottom two panels), in the midplane region $z=0$
and for the run PL20-512.  Surprisingly, we found that the largest
scale axisymmetric mode ($k_x\equiv k_{x_0}=2\pi/L_x, \,\,k_y=0$)
dominates
the energy content. This mode oscillates at a very well-defined
angular frequency $\omega_f=0.83\, \kappa$, close to the orbital or
epicyclic frequency. This is indicative of a large-scale inertial
oscillation modified by self-gravity. The mode is clearly responsible for the
pseudo-periodic variation of the Reynolds stress observed in Section
\ref{sec_prop}. Smaller scale axisymmetric modes such as $k_x=2k_{x_0}$
or $k_x=3k_{x_0}$ are negligible compared to the fundamental mode
(their specific energy decreases with their radial scale) and their
evolution appears more chaotic. 

Figure \ref{fig_energy}  (bottom panels) shows
that, energetically, individual spiral or non-axisymmetric waves grow
and decay aperiodically, but their amplitude remains on average weaker
than the largest scale axisymmetric mode 
(in particular $\ell=2$ at $t=42/\Omega$ or the orange and
blue ones at $t=57/\Omega$ and $t=60/\Omega$).
The energy of the
strongly amplified waves evolves quasi-exponentially during their
growth phase before soon decaying. 
Overall the weaker strength of spiral waves is surprising,
given that we expect gravito-turbulence to be primarily supported by
non-axisymmetric structures (axisymmetric modes being stable for
$Q_{\text{turb}}\geq 1$). It is important to note, however,
that we only have shown the waves' kinetic energy. 
The associated density perturbations associated with spiral waves
in fact can be larger than for axisymmetric 
 modes. Indeed, we find that
the density of the fundamental axisymmetric mode is 
rather small and does not oscillate regularly in time.

 \begin{figure}
\centering
\includegraphics[width=\columnwidth]{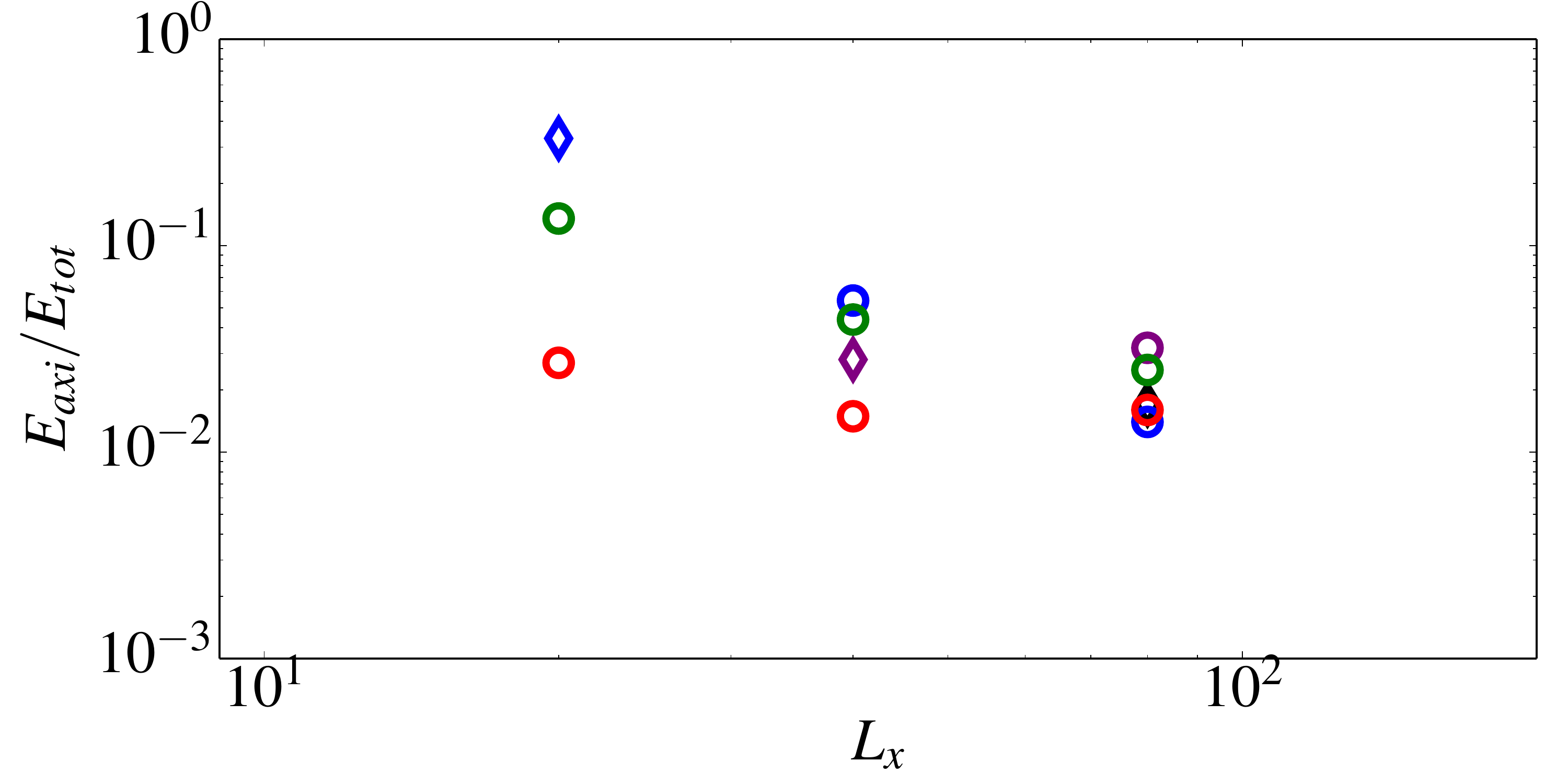}
 \caption{Ratio between the kinetic energy of  individual axisymmetric
   modes and the total kinetic energy (averaged over at least $100\,
   \Omega^{-1}$) as a function of horizontal box size $L_x=L_y$ (but
   same resolution per $H_0$). Black, purple, blue, green and red
   account for modes $k_x= 2\pi/80 j$,  $j=1,..,5$, in this
   order. The diamond marker indicates the fundamental mode with
   wavelength equal to the box size.}
\label{fig_aximode}
 \end{figure}
 
Are these axisymmetric modes physical or artificially
enhanced by the box size, or even the shearing periodic
 boundary conditions? To answer this
question, we determined the energetics of the axisymmetric
structures in larger boxes. Fig.\ \ref{fig_aximode} shows that when
$L_x=40$ and $L_x=80\, H_0$, the kinetic energy of each
individual axisymmetric mode is smaller than when $L_x=20 \, H_0$
but in total represents still $\sim 20 \%$ of the energy.  Also, for
$L_x \geq 40 \,H_0$, the kinetic energy associated with each
axisymmetric mode appears to converge to a similar value. Most
importantly, for a sufficiently large box, 
 the fundamental
mode (on the box size) ceases to be dominant, its place taken by a
shorter scale mode. Larger boxes can host multiple
axisymmetric inertial oscillations with frequencies close to $\kappa$.
These results
 indicate that the large-scale axisymmetric
dynamics is not controlled by the box size, which is consistent
with Section \ref{resolution} which showed some turbulent
quantities had converged
 by $L_x=L_y\gtrsim 40\, H_0$. 

Finally, we
checked that for large boxes, axisymmetric modes still exhibit an
oscillatory behaviour with a frequency close to $0.8 \kappa$  
(this in particular true for the harmonics $k_x=2 k_{x_0}$ or $k_x=3
k_{x_0}$). 
We checked also that the dynamics and amplitude of 
these modes are similar in both PLUTO and RODEO simulations.
In sum, oscillatory axisymmetric waves are an important component of
the GI  dynamics, regardless of the box size. Although
 their amplitude might be enhanced artificially if the box size is
 too small, their existence is likely to have a physical origin. 

\subsection{Origin and role of axisymmetric modes}

\begin{figure}
\centering
\includegraphics[width=\columnwidth, trim=0cm 16.5cm 0cm 0cm, clip=true]{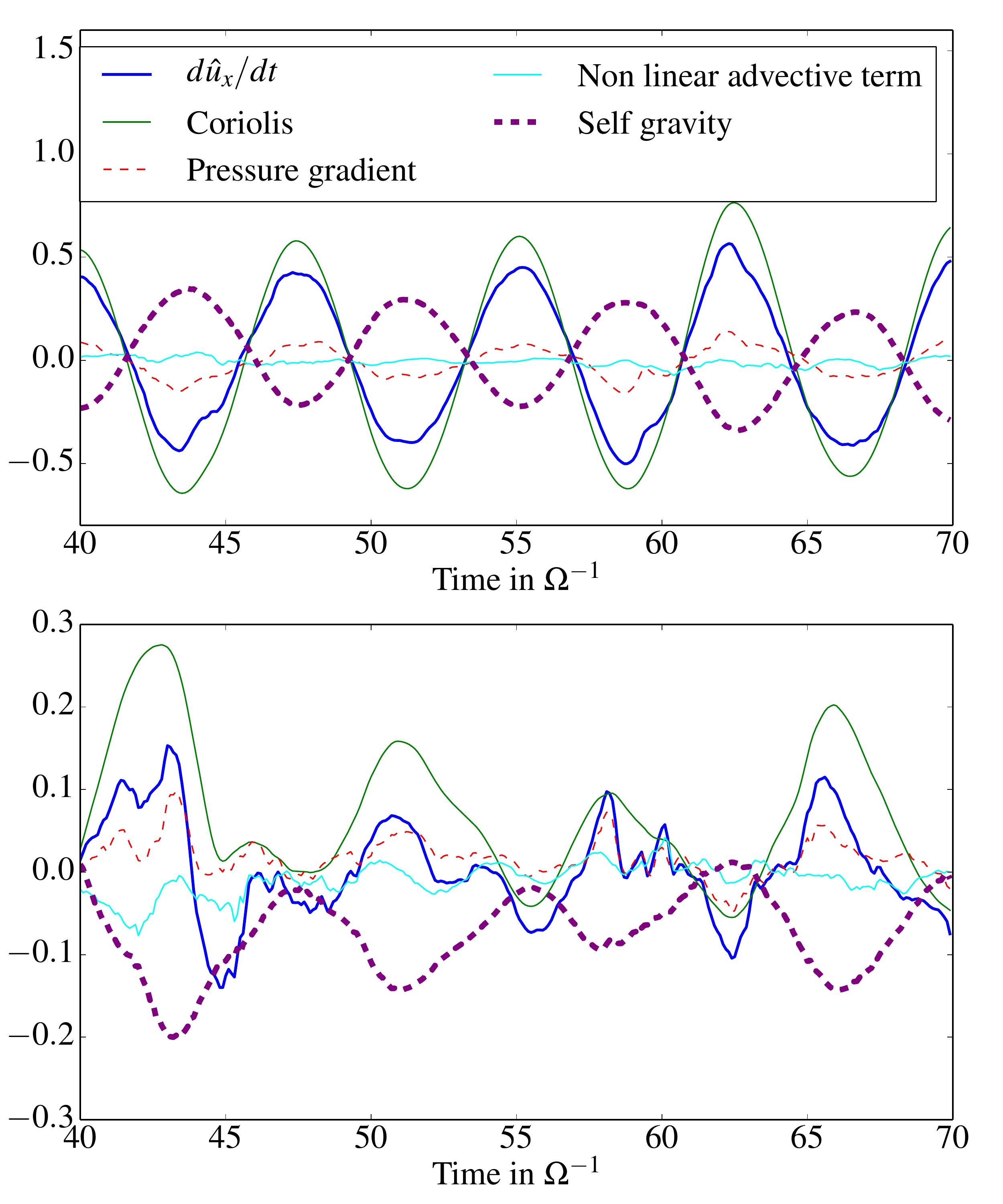}
 \caption{Radial force balance of the large scale axisymmetric mode $k_{x_0}=2\pi/L_x$ in the midplane region $z=0$.}
\label{fig_epicyclic}
 \end{figure}

Though non-axisymmetric shearing waves form the backbone of the nonlinear GI
dynamics, we have seen that large-scale axisymmetric 
oscillations are conspicuous participants. Several questions may then
be asked: 
\begin{enumerate}[(a)]
\item What is their origin? Though linearly stable, how can they reach
  and sustain large amplitudes?
\item What role do they play in the dynamics? Are they involved in the
  subcritical transition to, and the maintenance of, turbulence, or
  are they just passive modes? 
\end{enumerate}
In order to answer (a), we determine the force balance associated with
the fundamental axisymmetric mode $k_x=2\pi/L_x$ in PL20-512. It can
be decomposed into a high and low amplitude standing wave,
 $\hat{u}_1(t)\cos(k_xx)$ and $
\hat{u}_2(t)\sin(k_xx)$ respectively, with $\hat{u}_1 \gg \hat{u}_2$. 
Figure \ref{fig_epicyclic} shows that the first
wave possesses an inertial character since it is mainly driven by the
Coriolis force and partially restored by pressure. 
Self-gravity is always opposed to the latter and tends to
destabilize the wave (although $Q$ is too large for the wave to become
unstable). Nonlinear terms seem negligible in its dynamical
evolution, although they have a cumulative positive feedback on its energy budget. 
We checked that the force balance is similar in the azimuthal
direction (not shown here) and for the 
smaller amplitude second wave.  In sum, we identify a coherent
quasi-linear epicyclic oscillation. We found that this oscillation is
maintained for the entirety of our simulations (which is $3000\,
\Omega^{-1}$ for PL20-128b), and the
mode cannot be a vestige of our initial condition, since turbulent viscosity
would have dissipated it in a shorter timescale (typically $t_{\nu}
\simeq 1/(\alpha c_s \Omega k^2)\simeq 500\, \Omega^{-1}$).  This
suggests that large scale axisymmetric perturbations are nonlinearly
excited and regenerated by gravito-turbulence. 

 We performed several additional
simulations, shown in Appendix \ref{appc}, which strongly indicates
that the mode $k_x=k_{x_0}$ is the result of a nonlinear baroclinic interaction
between non-axisymmetric density waves. This dynamics
is reminiscent of the mode coupling proposed by
\citet{laughlin94} for gravitoturbulent discs and
by \citet{lee16} in the context of planet-disc interactions.
\citet{laughlin94} showed, in particular, that self-interaction of $m=2$
modes can lead to a nonlinear growth of the $m=0$ mode in global
discs. However their initial density profile is subject to a Rossby
wave instability, which complicates interpretation of their results.

Regarding question (b), there are several reasons to think that these modes
are active participants in the disc dynamics. First, they appear to be
sufficiently coherent in time to enter in resonance with a pair of
inertial waves. Such a resonance, based on a nonlinear triadic
interaction, is known to provide a path toward parametric
instabilities. We investigate this possibility in the next section.
  
  Second, the large scale axisymmetric modes could be involved in the
  re-generation and instability of leading shearing waves and thus, in
  the sustenance of the whole GI dynamics. A potentially related 
  regeneration
  process has been formulated for the 
 magneto-rotational dynamo problem \citep{Herault2011}.  
 Individual non-axisymmetric wave
  are sheared out into strongly trailing decaying structure. Long
  lived dynamics can only be excited if new leading waves are seeded
  each $\Delta t=\frac{2}{3} L_y/L_x \Omega^{-1}$. \citet{Herault2011}
  found that axisymmetric modes play the role of `walls' and thus 
  `confine' the non-axisymmetric
  waves.  The dying trailing structures can then be reflected
  and give birth to a new leading shearing wave. This mechanism is
  probably essential to any kind of sustained non-axisymmetric
  turbulence in the presence of a background shear flow.  
  Analogously, axisymmetric zonal flows are known to support a
  linear instability of self-gravitating spiral waves
  \citep{lithwick07,vanon16}. We suspect then that such an instability is at work
  in our simulations of gravito-turbulence and might feature
  prominently in the 
 self-sustaining process. The dynamics involving the nonlinear
  regeneration of axisymmetric modes (see previous paragraph and
  Appendix \ref{appc}) as well as the non-axisymmetric GI linear
 instability (supported by axisymmetric structures) might form the
 basis
  of the fully developed turbulence.  
  
\section{Excitation of a parametric instability}
\label{sec_parametric}

Section \ref{sec_fourier} revealed that a large-scale axisymmetric
epicyclic mode, oscillating at a frequency ${\omega_F} \simeq
0.8\, \kappa$, naturally emerges from gravitoturbulence. From
the perspective of the smaller scales, this
internal coherent oscillation may function equivalently to an external
periodic forcing or deformation of the disc. 
Rotating flows subject to periodic forcing or deformations are known to
excite parametric instability \citep{goodman93}; and in fact
the standing epicycle we witness in our simulations is 
similar in many respects to that treated
by \citet{fromang07c} who showed that
 an axisymmetric density wave
is subject to such an instability. In this
section, we explore this idea and consider whether a similar
instability occurs in our problem and is responsible
for the small scale turbulence 
appearing at $z\simeq H_0$. 

\subsection{Theoretical expectations}

\subsubsection{Onset of instability}

Parametric instabilities can arise when an oscillator
 undergoes a forcing at twice its natural frequency. In
accretion discs, they occur when 
a large-scale, time-periodic disturbance enters into resonance with a pair of
small-scale inertial waves \citep{goodman93}. This
fundamentally 3D instability has been studied exhaustively in the context of
eccentric \citep{ogilvie01b,papaloizou05,barker14} and warped discs
\citep{gammie00,ogilvie13}, and in the case of mean-motion resonances
between a disc and its companion \citep{lubow91}. More recently, it
has been shown to attack spiral density waves excited in the disc
 \citep{nelson16,bae16}. \\

Although relatively robust, parametric instabilities require certain
conditions to work.
If we denote by $\omega_{i_1}$ and $\omega_{i_2}$ the frequencies of 
two inertial waves and $\omega_F$ the frequency of
the large scale oscillation, 
it is possible to show that growth 
is obtained when \citep{fromang07c}:
\begin{equation}
\label{eq_param_condition}
\omega_F=\omega_{i_1}+\omega_{i_2}.
\end{equation}
To obtain a constraint on the background oscillation 
$\omega_F$, we need the dispersion
relation for each inertial wave. This is available 
in the WKBJ limit of short radial and
vertical wavelengths (Goodman 1993). We consider only the coupling of 
`neighbouring' inertial wave branches, for which $\omega_{i_1}\approx
\omega_{i_2}$, because of the density of
the spectrum on short scales. The resonance
condition immediately yields
\begin{equation}
\label{eq_wiwf}
\omega_{i_1}=\omega_{i_2}=\tfrac{1}{2}\omega_F,
\end{equation}
and because $\omega_{i_1}^2 < \kappa^2$, we deduce that 
 $\omega_F<2\kappa$. From the dispersion relation again, we 
find that resonance implies $\omega_{i_1}^2>N^2$, and so 
the condition
for the excitation of parametric instability is:
\begin{equation}
\label{eq_condition}
2\, N<\omega_{F}<2\, \kappa.
\end{equation}
\citep{goodman93,fromang07c}. 
As discussed by \citet{nelson16}, the influence of
buoyancy 
is considerable and must stabilize the flow above a 
certain disc altitude $z_{\text{crit}}$ for which $N>\kappa$.

We should emphasize that these simple theoretical arguments have their limits.
First, condition (\ref{eq_condition}) is derived for axisymmetric
disturbances, though for weakly non-axisymmetric waves $(k_y \ll k)$
\citet{goodman93} showed that condition (\ref{eq_condition}) still
holds. Unlike axisymmetric waves, however, their lifetime is finite
and they can only be amplified transiently.  Second, condition
(\ref{eq_condition}) only applies to localised wave packets, 
and not to vertically global modes. 
In principle one could solve the full linear eigenvalue problem 
with vertical structure, similarly to \citet{ogilvie13} and \citet{barker14}, but we leave this task
for another time. In any case, given the disordered background in
which the
inertial waves find themselves, they may manifest more as localised
packets rather than larger-scale organised structures. Finally,
 self-gravity has been neglected, though for $Q\simeq 1$ and $k_x$ large, the dispersion
relation of linear inertial disturbance is almost unchanged when compared
to the case without self-gravity (see  Fig.~3 in \citet{mamat10}). 

In the isothermal case, \citet{fromang07c} estimated the growth rate
of parametric 
instability due to an axisymmetric density wave to be 
roughly proportional to the product of the
fractional amplitude of the density wave and $\omega_F$. Applying this
formula to the large-scale epicyclic oscillations in our simulations this
gives a growth rate $\sim 0.1 - 0.2 \,\Omega$, and so we expect the parametric
instability to develop after few orbits in our simulations, unless
it is impeded by other participants in the gravitoturbulence.

\subsubsection{Propagation and breaking of inertial waves}
\label{inertial_breaking}

Suppose that the conditions for exciting inertial
waves are met. How do these waves
propagate and dissipate in the disc? First note that,
as a localised wave packet
travels
upwards from the midplane, the background state around it changes
and, as a consequence, the wave packet properties will also change. In fact,
the wave will `refract' and become `channeled' by the disk's
vertical
structure, which acts similarly to a wave-guide \citep{lubow93,
  bate02}. The packet's upward trajectory will bend more and more
radially, before
ultimately settling into a normal eigenmode of the vertically 
global problem, travelling solely in the radial direction. 

Perhaps before that occurs, the wave will break nonlinearly. 
In an unstratified medium,
inertial waves are known to break down into turbulence
via the action of secondary instabilities, driven by shear,
which occurs when displacements are comparable to
wavelengths. If the inertial wave has a form $\hat{u} \exp{\text{i}(k_\xi \xi -\omega
  t)}$ where $\xi$ is in the  
direction of propagation, then breaking occurs when the
particle velocities exceed the phase velocity
\begin{equation}
\hat{u} \approx \omega /k_\xi.
\end{equation}
This relation is equivalent to saying the Rayleigh stability
criterion is locally violated (the angular momentum gradient becomes
locally negative, see \citet{wienker16}). This secondary instability
is actually then another parametric instability involving a primary
inertial wave and two daughter waves. The route to turbulence in such
systems may then be viewed as a cascade of parametric instabilities. 

In a stratified disc, inertial waves  naturally break as they propagate
toward the surface of the disc. Indeed, since the density decreases
above the midplane, the velocity components increase with $z$
(consider the linear
eigenfunctions of a stratified disk, Appendix \ref{appA}). We expect
inertial waves to break above a critical height
that depends on the mode structure. For instance, a $n=1$ inertial
mode, for which
$\hat{u}_x \propto z$ (see Appendix \ref{appA}), possesses a critical
 height around $z \simeq
H_0$. Higher order $n$ modes break at altitudes somewhat lower, but
all are consistent with our observation of small-scale activity above
the midplane. Most of the kinetic energy transfers to small scale at
this critical height or above. Energy is then expected to saturate at a value of order $u^2=\omega^2 /k_\xi$, although a more complete calculation,
in the context of eccentric discs, shows that it also increases with the
amplitude of the parent wave \citet{wienker16}.

\subsection{Evidence of parametric instability from the numerical data}

\subsubsection{Temporal spectra}

 \begin{figure}
\centering
\includegraphics[width=1\columnwidth]{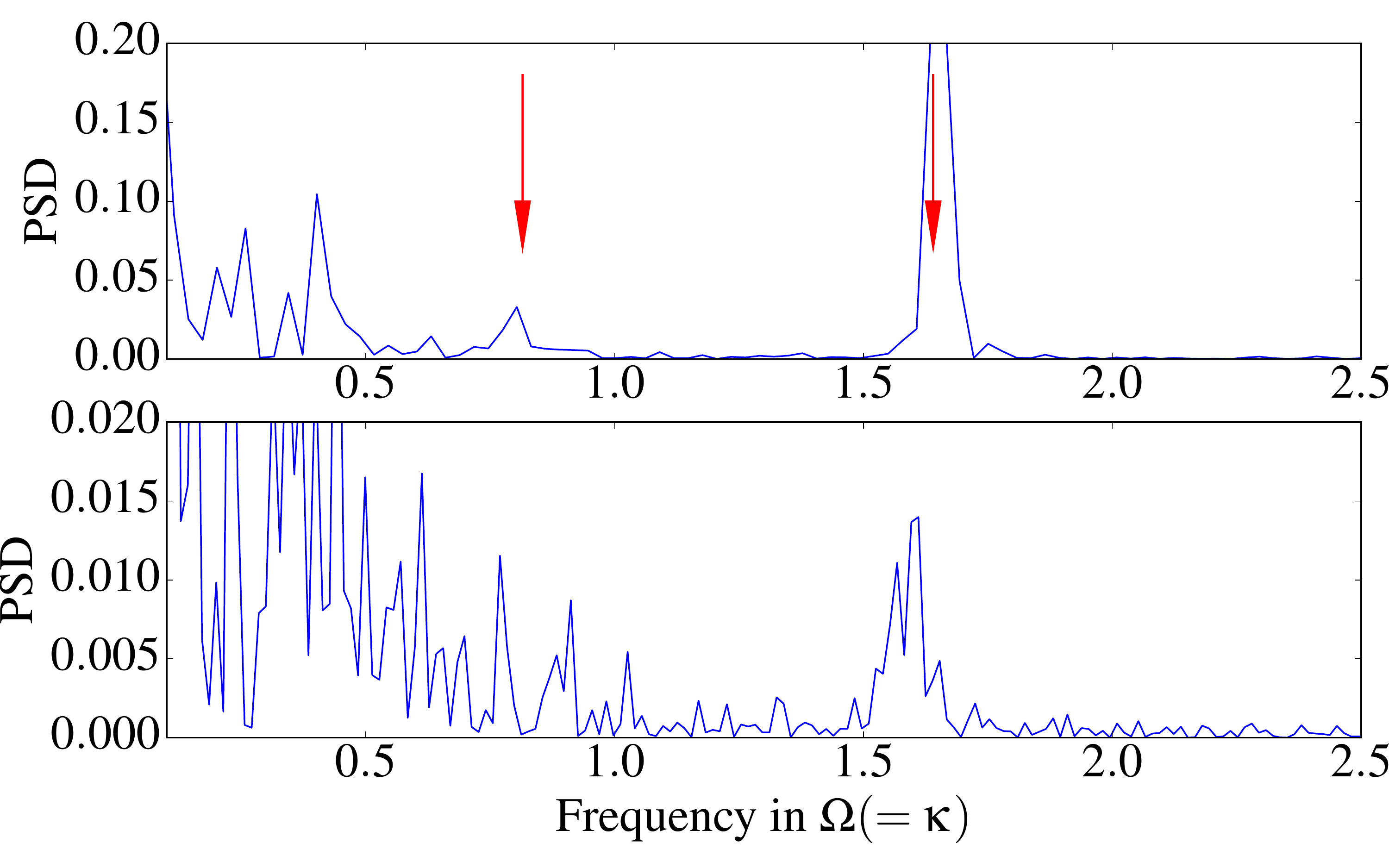}
 \caption{Power spectral density (PSD) of the averaged kinetic energy
   as a function of temporal frequencies. The upper panel 
   was obtained from simulations
   PL20-512 and the lower panel from PL20-128. The red arrows on the
   right indicate the frequency of the axisymmetric
   oscillation (note that the kinetic energy oscillates at twice the natural
   frequency of the mode). The left arrows show a secondary  
   peak at 0.8 $\kappa$,  which might correspond to the resonant inertial modes. }
\label{fig_tfft}
 \end{figure}

In Section \ref{sec_fourier} we identified a large scale axisymmetric
mode that oscillates at a frequency $\omega_F\approx
0.8\kappa$. According to condition \eqref{eq_condition}, a parametric
instability is then possible in the vicinity of the midplane where
$N=0$ and should induce the growth of a pair of inertial waves at a 
frequency $\omega=\omega_F/2$.
To assess whether or not the instability occurs in our simulations, 
we compute the temporal Fourier spectrum of the
turbulent signal. Fig \ref{fig_tfft} (top) shows the power spectrum of
the box-averaged kinetic energy as a function of (temporal) frequencies
for the PL20-512 run. A strong peak is observed at a frequency
$2\omega_F$, which corresponds to the large-scale axisymmetric mode
(quadratic quantities like energy oscillate at twice
the frequency of the associated mode). From the right to the left, we
see a second peak, smaller in amplitude, at half the frequency of the
fundamental peak $\omega=\omega_F$. If we look further to the left, several
distinct peaks are seen respectively at $\omega=\omega_F/2$ and
$\omega=\omega_F/4$. This cascade of subharmonics peaks at frequency
$\omega_F/2^n$  may be interpreted as a direct signature of parametric
instability via inertial mode couplings. 

A similar result is obtained
from the PSD of PL20-128, calculated over a much longer time $3000 \, \Omega^{-1}$, although
the signal is more noisy and the peaks less pronounced. Because of the
rich assortment of long-time variability (discussed in Section 3.1) it
is difficult to extract as clear a signature of the parasitic inertial
waves, as in the shorter time high-resolution run
PL20-512. 

In a 
larger boxes with $L_x=40 \, \Omega^{-1}$ and $L_x=80 \, \Omega^{-1}$,
the signal is spread over more frequencies and the spectrum is
more difficult to interpret. Now \emph{multiple} large-scale axisymmetric
modes are hosted in the box with similar frequencies (cf.\
Fig.~\ref{fig_aximode}),
each potentially
unstable to different parametric resonances. The ensuing turbulent
dynamics is more complicated and the temporal spectrum a less useful
diagnostic in this case.

 \begin{figure*}
\centering
\includegraphics[width=1\textwidth]{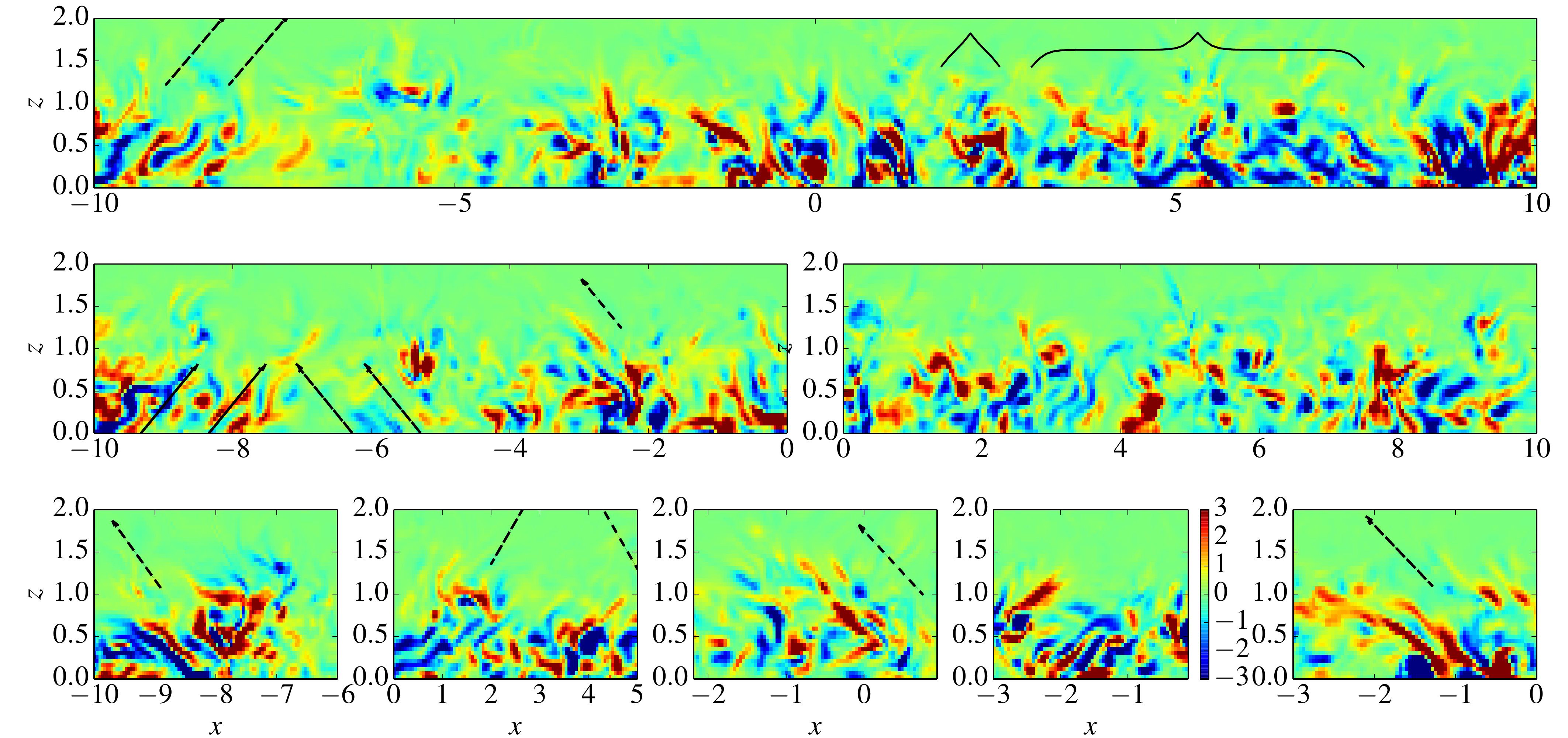}
 \caption{Helicity in the poloidal $(x,z)$ plane weighted by the
   average density profile in $z$, for different time. All arrows
   indicate a direction of $45 \degree$ with respect to the vertical
   axis. The curly brackets shows some examples of intricate and
   tangled patterns probably associated with inertial waves. In any of these figures, it is possible to see an alternation of inclined red and blue bands, with typical wavelength $\approx 0.1 - 0.2 H_0$}
\label{fig_helicity}
\end{figure*}
\begin{figure*}
\centering
\includegraphics[width=0.87\textwidth]{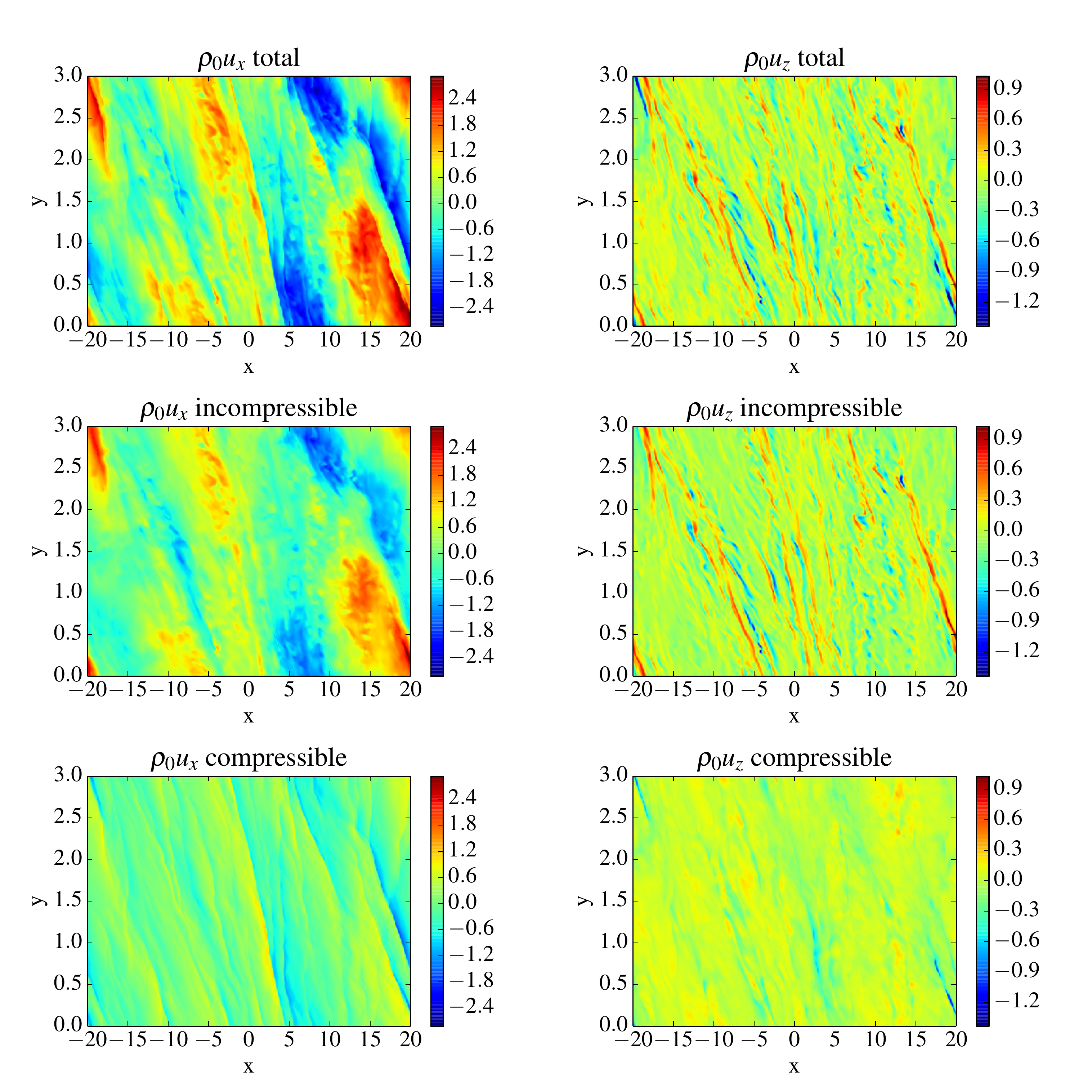}
 \caption{Velocity components $u_x$ (left) and $u_z$ (right) in the mid plane $z=0$, weighted by the average density at that location. The top panels represent the total field, the center and bottom panels represent respectively its incompressible and compressible parts.}
\label{fig_hdecomp1}
 \end{figure*}
 \begin{figure*}
\centering
\includegraphics[width=0.87\textwidth]{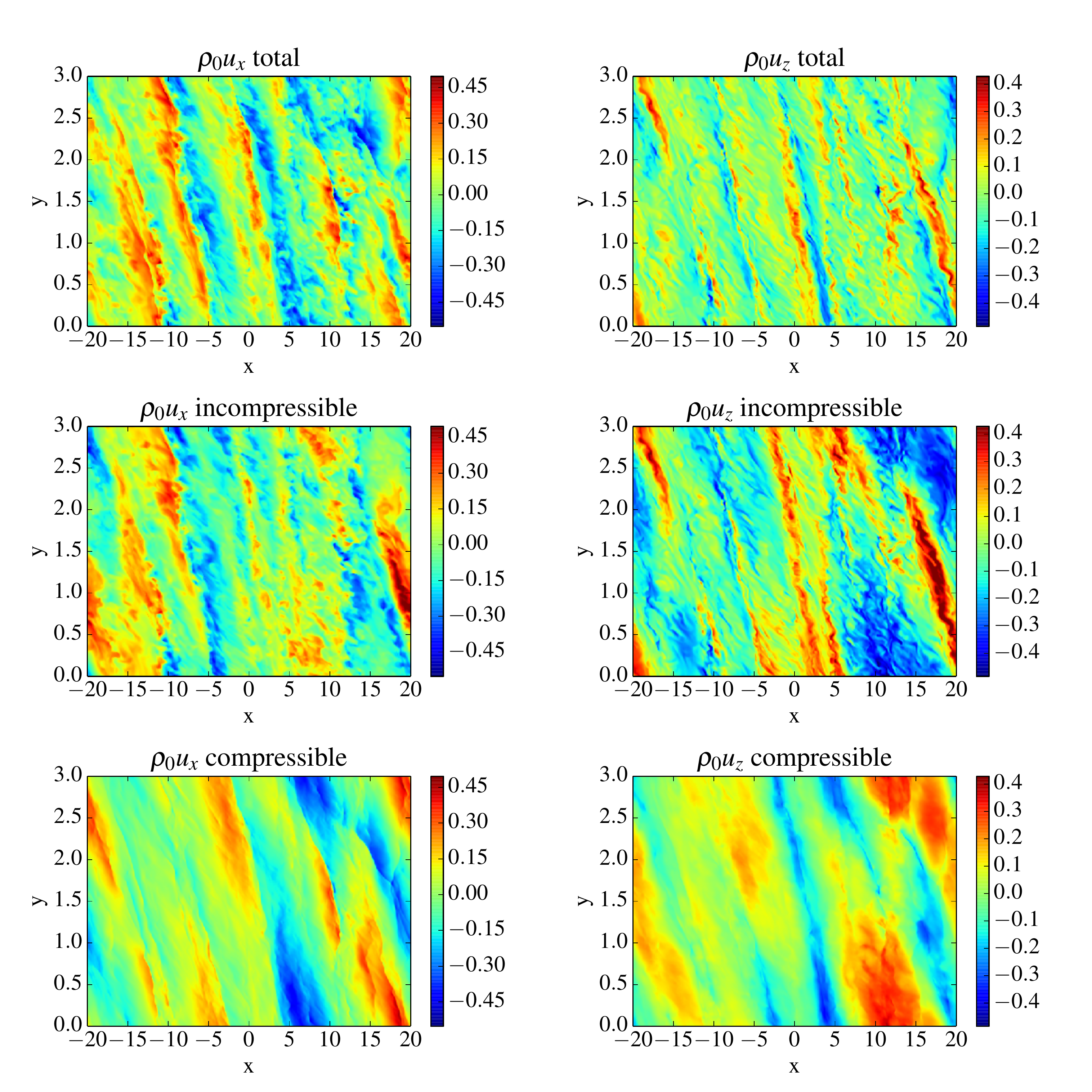}
\caption{Velocity components $u_x$ (left) and $u_z$ (right) in the plane $z=H_0$, weighted by the average density at that location. The top panels represent the total field, the center and bottom panels represent respectively its incompressible and compressible parts.}
\label{fig_hdecomp2}
 \end{figure*}

\subsubsection{Helicity maps}

Another way to trace the inertial waves and the associated 
parametric instability is to compute the helicity of the flow 
\begin{equation}
\mathcal{H}=\mathbf{u}\cdot (\mathbf{\nabla}\times\mathbf{u}).
\end{equation}
Because inertial waves satisfy
$\mathbf{\nabla}\times\mathbf{u} \simeq \pm k \mathbf{u}$, where $k$ is
their wavenumber, they are intrinsically helical. The direction
of propagation of a given packet corresponds to a surface of
constant helicity. These different properties have been studied in
particular in the context of the solar and geodynamo
\citep{ranjan14,davidson14,wei15,singer83}.

Figure \ref{fig_helicity} shows the helicity $\mathcal{H}$, weighted by
the averaged vertical density profile, projected onto the $(x,z)$ plane
at different $y$ and times. Helical structures have a typical size of
$\lesssim H$, in both the $z$ and $x$ directions. They form inclined
and interleaved layers of positive and negative helicity,
corresponding to the troughs and crests of inertial waves. Thus the
associated wavevectors cut across the layers, while the group velocities
point along the layers (Bordes et al.~2012). In
Fig.~\ref{fig_helicity} the inclined black arrows have been
superimposed to highlight the inertial layers.

The preferential angle of
propagation is in a range between $\pm 30\degree$ and $\pm 50\degree$
measured from the vertical axis. However, using the local, linear and axisymmetric dispersion
for inertial waves \citep{goodman93}, and setting $\omega_i =
\frac{1}{2}\Omega_F\approx 0.42\kappa$, we
find that unstable inertial waves should possess a group velocity
pointing within a range of angles  $ \lesssim 30\degree$ with respect to the
rotation axis. The slight discrepancy with the simulation data we
attribute to wave channelling by the disk structure, as discussed
earlier. In fact, the bottom right panel of Fig.~\ref{fig_helicity}
presents a nice example of wave channeling.

\subsubsection{Helmholtz decomposition and connection with the small-scale turbulence}

We
showed in Sections \ref{resolution} and \ref{smallscaledyn}
vigorous small-scale turbulent activity taking place $z \sim H_0$, 
while in the previous subsection we compiled numerical 
evidence for inertial waves propagating upward at altitudes less than
$H_0$. In this subsection we attempt to connect the two.

First we check that the small-scale motions are not directly excited by
large-$k$ gravitational modes. We performed a simulation, starting
from a gravito-turbulent state, taken from PL20-512, and filtered the
gravitational potential on small-scales by retaining only modes
with $k_x<6k_{x_0}$ and $k_y<6k_{y_0}$. Despite the filtering, 
the small-scale turbulence persisted for long times in the 
reconfigured simulation. 

This activity, however, might still consist of a field of small-scale
p-modes, different in character to inertial waves.
In fact, short wavelength linear p-modes
preferentially localise at the disk surface \citep{kory95,ogilvie98waves}, where the turbulence is observed.  They may be generated by disordered motions on large-scales, in particular by the nonlinearly
breaking of large-scale wakes, which may favour the less-dense upper layers.

One way to disentangle these different small-scale motions
is to decompose the velocity
field into compressible and incompressible parts. Inertial waves
are generally incompressible (or rather anelastic) 
if they vary on a vertical scale comparable or less than the
background support ($\bar{k}_z \lesssim 1$). On the other hand,
p-modes are associated with compressible motions,
though they are not necessarily curl-free and can also possess an incompressible part.
To separate the compressible part of a vector field
 $\mathbf{F}$ from its incompressible part, we use the Helmholtz decomposition:
\begin{equation}
\label{eq_helmholtz}
\mathbf{F}=\mathbf{F}_{\text{c}}+\mathbf{F}_{\text{ic}}=\mathbf{\nabla}\varphi + \mathbf{\nabla}\times\mathbf{{\Phi}},
\end{equation}
where $\varphi$ is a scalar field, defined up to a constant and
$\mathbf{\Phi}$ a vector field defined up to a gradient field. The
first term is the compressible and potential part of
$\mathbf{F}$. This is a curl free component, i.e
$\mathbf{\nabla}\times\mathbf{F}_{\text{c}}=0$. The second term is the
incompressible and solenoidal part, and it 
is divergence free, i.e $\mathbf{\nabla}\cdot\mathbf{F}_{\text{ic}}=0$.
Applying the divergence operator on both sides of Eq.~\eqref{eq_helmholtz} and using the solenoidal properties, we obtain
\begin{equation}
\nabla^2\varphi = \mathbf{\nabla}\cdot\mathbf{F}.
\end{equation}
Therefore, to find the potential $\varphi$, we have to solve a Poisson
equation with a source term $\mathbf{\nabla}\cdot\mathbf{F}$. In the
shearing box frame, the method is very similar to the one used to
obtain
the disk's potential (see Section
\ref{poisson_solver}), so we simply adapted our Poisson solver to deal 
 with a different source term. Several tests on simple flows were
 completed
 before applying our method to the full turbulent field.
The calculation of $\mathbf{\Phi}$ requires solving  three
Poisson equations but in practice we obtain 
it by substracting the compressible part from the initial field $\mathbf{F}$. 

We applied the Helmholtz decomposition to the vector
$\mathbf{F}=\rho_0\, \mathbf{u}$, where $\rho_0$ is the vertical profile
of the horizontally-averaged density. This decomposition permits us
to analyse the incompressibility of the flow, remembering that
 the incompressible component 
$(\rho_0\, \mathbf{u})_{ic}=\mathbf{\nabla}\times\mathbf{{\Phi}}$ will
 preferentially exhibit an inertial character.

Figure \ref{fig_hdecomp1}
shows the incompressible (centre panels) and compressible parts (bottom
panels) of the flow components $\rho_0\, u_x$ and $\rho_0\, u_z$ in
the midplane $z=0$. For the radial component $\rho_0\, u_x$, the two
parts have similar amplitudes and
both exhibit large-scale spiral waves. In the incompressible
part the waves manifest as large-scale features, while in
the compressible part they are visible as thin shocks. In
addition, we observe small-scales bundles, of typical size
0.2 $H_0$, in the incompressible part exclusively. The vertical
velocity (right panels) is clearly dominated by the
incompressible component, which exhibits 
thin filament structures organised along the
wakes. We attribute both the small vertical velocity filaments and the
horizontal bundles with incompressible inertial waves developing
at the midplane. 

Figure \ref{fig_hdecomp2} shows
the same decomposition at $z=H_0$. Here the small-scale dynamics is
far more developed, in both the radial and vertical velocities. 
Most importantly, however, is that these small-scale features are
almost completely confined to the incompressible parts of both
velocity components. The compressible parts exhibit predominantly
large-scale structure associated with the gravity wakes, in the case
of $u_z$ arising probably from 
the vertical breathing or splashing of the wakes. There is
no evidence of small-scale p-modes as might be generated by wave
breaking at this altitude. In summary, the marked incompressibility of
small-scale turbulence at $z=H_0$ is strong evidence that it is indeed
comprised of inertial waves excited by parametric instability,
breaking and mixing at these altitudes 
(as described in Section \ref{inertial_breaking}).

\label{coolingfrag}
\section{Conclusions and astrophysical implications}
\label{sec_discussion}

We performed a set of 3D vertically stratified shearing-box simulations of
gravitoturbulence, reproducing the local behaviour of marginally
gravitationally unstable discs. For a fixed cooling time
$\tau_c=20\,\Omega^{-1}$, we provided evidence showing that averaged
turbulent properties
converge for box size larger than $40 \,H_0$ and for sufficient
resolution. The convergence of some quantities, however, are difficult
to ascertain, such as the kinetic energy and Reynolds stress, due to
their bursty and very long time variation (hundreds of orbits).

Our highest resolution runs exhibit small-scale disordered
motions at around $z\approx H_0$.
It is unclear whether the properties of this small-scale turbulence have yet converged.
 We compile theoretical and numerical evidence that the origin of the
turbulence is a parametric instability involving a pair of inertial
waves and a large-scale axisymmetric epicycle emerging naturally
from the gravitoturbulence. It is likely that this large-scale mode is
an important participant in the onset and sustenance of nonlinear
non-axisymmetric GI. 

We discount alternative causes of the
turbulence.
Kelvin-Helmholtz instability arising from the gravity 
wakes' horizontal shearing should lead to activity at all altitudes,
not just at $z=H_0$. Nonlinear breaking of the gravity wakes should
lead to compressible motions, but we show that the small-scale
turbulence is primarily incompressible at $z=H_0$. 
 Finally, vertical splashing due to
wake collisions should give rise to larger-scale motions dominated by
the vertical velocities, which is numerically observed only in the
compressible part of the flow. The small-scale turbulence we find is
incompressible and not dominated by vertical motions.

It should be stated clearly that our simulations employ large
horizontal domains ($>20 H$) for which it is challenging to justify
the locality of the shearing box model. This is
certainly the case for the thicker PP disks, where $H/r\sim 0.05$, but
there is more lee-way with AGN. We persist with this numerical tool
because
it affords us a well-defined and relatively clean platform
to probe shorter-scale dynamics at high resolution. Previous grid-based global
simulations can compete on vertical resolution, but suffer from poor 
azimuthal resolution especially \citep[e.g.][]{michael12,steiman13}. The radial boundary conditions in global
simulations impose
further complications that muddy their interpretation. 

There are several astrophysical implications of this work. We
focus on those most relevant to PP disks. 
Small-scale turbulence disturbs the coherence of the 
large-scale spiral waves, while not exactly destroying them. This
`scrambling' may be possible to observe in direct
imaging of large-scale spiral arms, giving them a somewhat
`flocculent' appearance \citep[see discussion in ][]{bae16}.

Perhaps of greatest importance is the impact of small-scale turbulence
on marginally coupled particles off the midplane. Two-dimensional planar
simulations of GI and dust indicate that this class of particle can collect in
filaments, sometimes achieving overdensities two orders of magnitude
over the background. Moreover, their relative velocities are
diminished to below 1\% of the sound speed, facilitating gravitational
collapse \citep{gibbons12, gibbons14, gibbons15, shi16}.
If, however, within these filaments there is an additional source of
turbulent motions, on a similar scale (as we find), then such overdensities may be
difficult to attain, and gravitational collapse more
difficult. Moreover, because the eddy-sizes of the parasitic
turbulence 
are $\ll H$ it is more likely that
pre-collisional pairs are decorrelated and will thus have large impact
speeds. However, for these effects to be significant the vertical
thickness of the particle subdisk must be $\lesssim H$, and this may not always
be a situation that lasts very long in the coupled evolution of the
disk and dust. 

More generally, small-scale turbulence will enhance whatever diffusion
is present. Figure \ref{fig_rms} shows that
better resolved turbulence exhibit larger rms velocities by factors
$\sim 1$, and one would expect diffusivities to be similarly
enhanced. 
As mentioned earlier, even our highest resolved
simulations are still probably unconverged with respect to this
physics. 

Though not explored in this paper, small-scale motions may make
fragmentation
more difficult. It should be emphasised that the typical
inertial wave motions
may be too slow to be effective, and being localised to $z\sim H$
will certainly be inefficient at the midplane where most fragmentation begins.
Work specifically targeting fragmentation in high resolution 3D disk
models will appear in a separate paper.

Finally, one may ask how these features affect the
magnetohydrodynamics in PP and AGN disks, as both may be moderately
well-ionised in their outer regions  \citep{menou01,turner14}. The preponderance of helical motions automatically
provide a means of dynamo action, but the magnetorotational instability (in some non-ideal form)
must also play a role here. One may then imagine a rich and complex
interplay between the GI, the parasitic turbulence feeding off it, and
the magnetorotational instability. 

\section*{Acknowledgements}

The authors thank the reviewer for a set of helpful comments,
and Richard Booth, Charles Gammie, Sebastien Fromang, and Ji-Ming Shi
for important feedback on an earlier draft of the paper.  
This research is
partially funded by STFC grant ST/L000636/1. Many of the simulations were run on the DiRAC Complexity system, operated by the
University of Leicester IT Services, which forms part of the STFC DiRAC HPC Facility (www.dirac.ac.uk). This equipment is funded
by BIS National E- Infrastructure capital grant ST/K000373/1 and STFC DiRAC Operations grant ST/K0003259/1. DiRAC is part of the UK National E-Infrastructure.

\bibliographystyle{mnras}
\bibliography{refs} 
\appendix

\section{Background equilibrium}

We present in this appendix several tests to verify that our
self-gravity module in PLUTO and boundary conditions are correctly
implemented in 3D. Note that the 2D linear problem (both 
axisymmetric and non-axisymmetric) has been already checked in \citet{riols16a}.
\label{appA}
In the presence of self-gravity the vertical background
equilibrium is governed by Eqs.~\eqref{eq_rho_eq} and
\eqref{eq_phi_eq},  where $c_{s_0}$ and $\rho_0$ denote the sound speed and the density in the midplane.
By introducing the `isothermal' Toomre parameter
$$Q_{{2D}_{0}}=\dfrac{c_{s_0}\Omega}{\pi G\Sigma},$$  and the
dimensionless $\bar{z}=z/H_0$, $\bar{\rho}=\rho/\rho_0$ as well as the
ratio $\Delta=4H_0\rho_0/\Sigma$,  the set of equation
(\ref{eq_rho_eq})-(\ref{eq_phi_eq}) reduce 
to a single dimensionless equation: 
\begin{equation}
\label{eq_eq_nd}
\dfrac{1}{\gamma}\dfrac{d}{d\bar{z}}
\left[\dfrac{1}{\bar{\rho}}\dfrac{d\bar{\rho}^\gamma}{d\bar{z}}\right]+1+\dfrac{\Delta}{Q_{{2D}_{0}}}\bar{\rho}=0.
\end{equation}

We numerically solve this equation by use of a finite difference
method. We impose $\Sigma=1$ and a certain value for
$Q_{{2D}_{0}}$. Imposing these two quantities makes the calculation
somewhat tricky because the midplane density $\rho_0$, and hence
$\Delta$, are not known in advance. It is then necessary to begin with
a guess for $\rho_0$ and to refine 
the solution iteratively, using a shooting method, until the surface density matches our prescription. 
 
Note that the isothermal Toomre parameter defined here is not exactly
equal to the $Q$ in the simulations, as the density-weighted average $c_s$
differs from the midplane one. However the sound
speed profile associated with the equilibrium scales as $\rho^{1/3}$
with $\gamma=5/3$, so that the difference between $Q_{{2D}_{0}}$ and
$Q$ defined in Section \ref{alpha} is small.  

\begin{figure}
\centering
\includegraphics[width=\columnwidth]{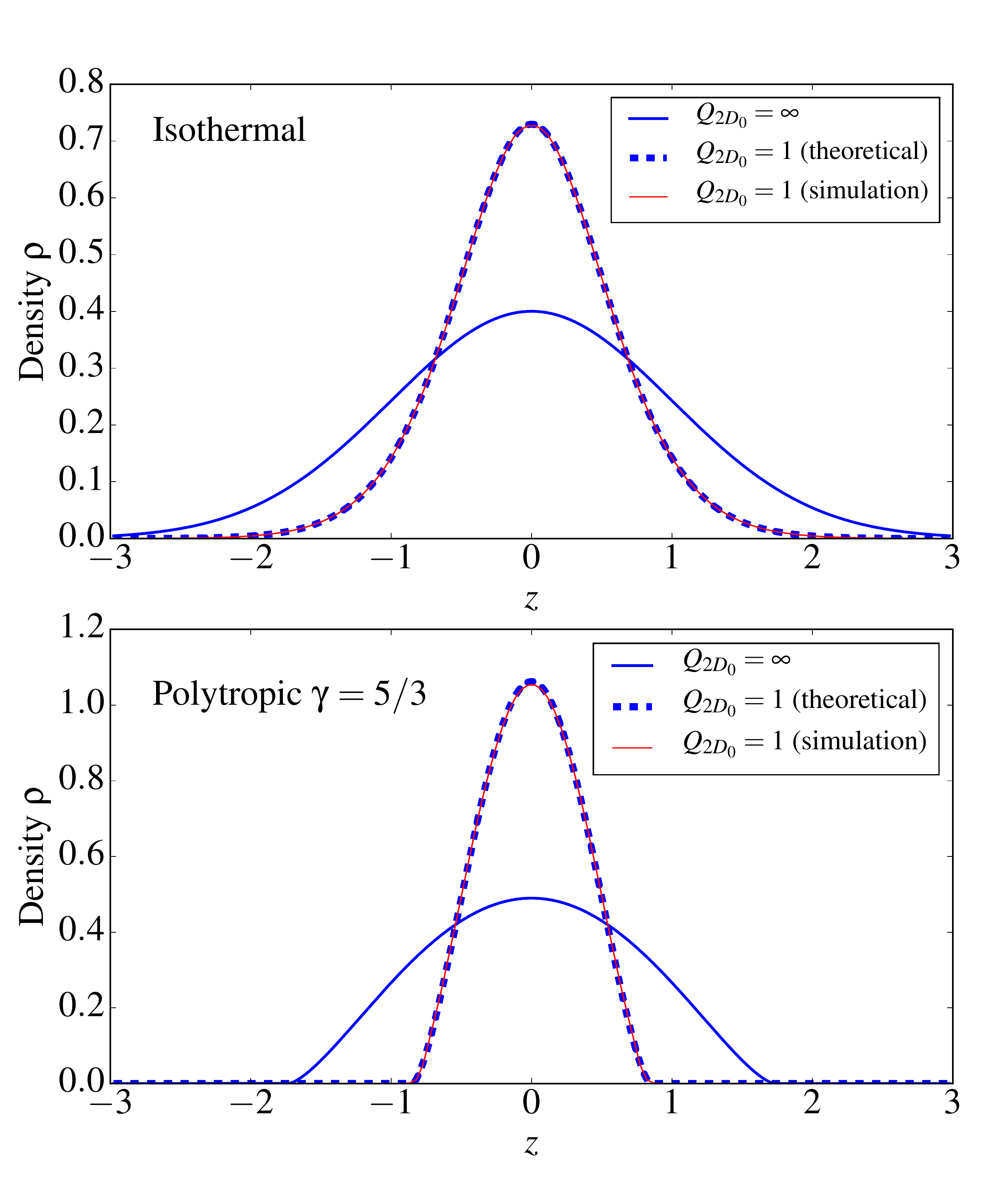}
 \caption{Density profiles of disc equilibria, for an
   isothermal gas (top) and polytrope with $\gamma=5/3$ (bottom)
   (fixed $\Sigma=1$). The solid blue line corresponds to the
   theoretical profile computed from Equation \eqref{eq_eq_nd} without
   self-gravity ($Q_{{2D}_{0}}=\infty$) while the dashed thick blue
   line is computed for $Q_{{2D}_{0}}=1$. In each 
 cases, the thin red line is the density profile obtained 
with PLUTO after $100\,\Omega^{-1}$ and is almost indiscernible from the theoretical one.}
\label{fig_rhoprofile1}
 \end{figure}
 \begin{figure}
\centering
\includegraphics[width=\columnwidth]{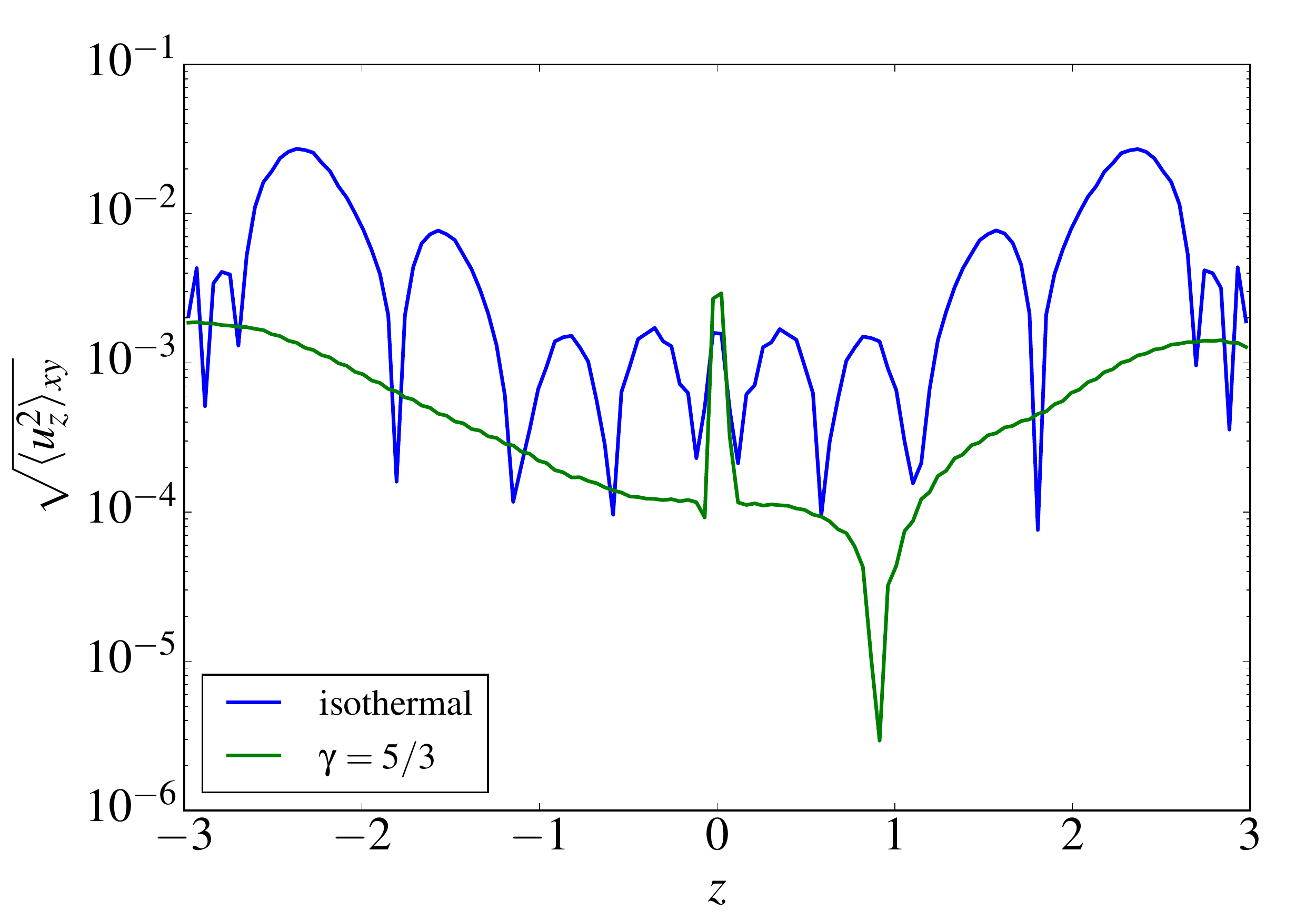}
 \caption{Vertical 
 profile of the r.m.s vertical velocity fluctuations at $t=100 \Omega^{-1}$ excited by vertical boundary conditions. }
\label{fig_rhoprofile2}
 \end{figure}

Figure~\ref{fig_rhoprofile1} shows different density (and pressure)
profiles for an
isothermal gas ($\gamma=1$) and a polytropic gas with $\gamma=5/3$,
obtained by numerical integration of Eq.~\ref{eq_eq_nd}.  When
self-gravity is included and $Q$ is of order 1, the disc becomes more
compressed and its effective height, defined such that
$\rho=e^{-0.5}\rho_0$, is $H_{\text{eff}}=0.52 H_0$  for the isothermal
case and $H_{\text{eff}}=0.41 H_0$ 
for the polytrope.

For $Q_{{2D}_{0}}=1$, we use these semi-analytical density profiles
(and the associated pressure) as an initial condition in PLUTO. We ran the
code for $100\, \Omega^{-1}$ and checked that the density has not
evolved during that  time (see the red solid lines in
Fig.~\ref{fig_rhoprofile1}). We  also plot in
Fig.~\ref{fig_rhoprofile2} the vertical  profile of velocity fluctuations at $t=100 \,\Omega^{-1}$. 
For stable configurations, the fluctuation amplitude
remains smaller than $10^{-2} c_{s_0}$. For a
polytrope, these  fluctuations can be larger than the local
sound speed, but do not influence the equilibrium in the midplane.
This shows that the boundary conditions are correctly implemented and
do not  introduce too much noise, even for the extreme case of a
polytrope for  which the density is floored to a certain value in the upper atmosphere. 

\section{Linear axisymmetric modes}
\label{appB}

\begin{figure}
\centering
\includegraphics[width=0.9\columnwidth]{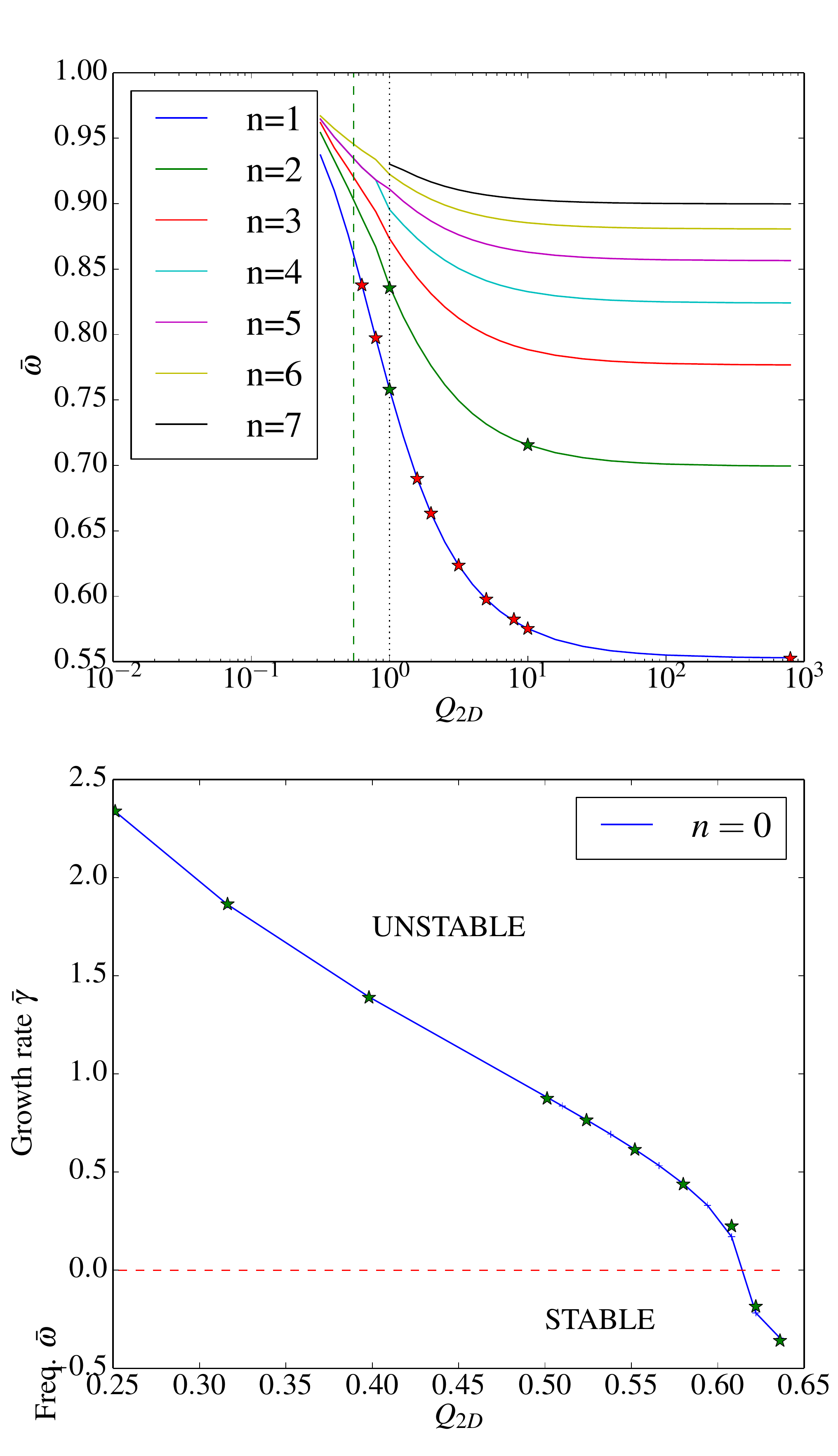} 
\caption{Properties of linear axisymmetric modes in the local
  isothermal thin disc approximation. Top panel: 
   dependence of the wave frequency $\omega$ on $Q_{{2D}_{0}}$ for the inertial
   branch modes $n=1,2,...,7$. Bottom panel: growth rates of the
   first unstable $n=0$ mode. In all cases $\bar{k}_x=2\pi/5$. Solid curves 
  are theoretical growth rates or frequencies, calculated with the
  linear eigensolver.
   Red and green stars are the values obtained from PLUTO simulations,
   with resolution of 256 and 512 points per azimuthal wavelength, respectively. The dotted line is $Q_{2D}=1$ while the dashed line represents the critical $Q_{2D}$ for which the $n=0$ mode becomes unstable.}
\label{fig_linearstab}
\end{figure}

To further test our Poisson solver in PLUTO, we simulate the
linear solutions and compare them with semi-analytic calculations. Unlike
the 2D case, it is not possible to analytically derive a
dispersion relation and we need to solve numerically the full
eigenvalue problem. We restrict our analysis to the isothermal case
and assume axisymmetric density and velocity perturbations of the form
$\left[\hat{\rho}(z),\,\hat{\mathbf{u}}(z)\right]\,\exp(\text{i}k_xx-\text{i}\omega
t)$.  We denote by ${\rho}={\rho}_e(z)$  and $\Phi_e(z)$ the
density and potential equilibrium, respectively. We normalize densities
(background and perturbations) by the midplane density $\rho_0$,
velocities by the uniform sound speed $c_{s_0}$, and pressure by $\rho
c_{s_0}^2$. Finally by introducing the dimensionless wave  
frequency $\bar{\omega}=\omega/\Omega$ and wavelength
$\bar{k}_x=k_xH_0$,
 the linearized and normalized system of equations reads
\begin{equation}
-\text{i}\bar{\omega} \hat{\rho}+\text{i}\bar{k}_x\rho_e \hat{u}_x+ \dfrac{d}{d\bar{z}}(\rho_e \hat{u}_z)=0
\end{equation}
\begin{equation}
\rho_e(-\text{i}\bar{\omega} \hat{u}_x-2 \hat{u}_y)=-\text{i}\bar{k}_x \hat{P}-\text{i}\bar{k}_x \rho_e\hat{\Phi}
\end{equation}
\begin{equation}
\rho_e(-\text{i}\bar{\omega} \hat{u}_y+\dfrac{1}{2}\hat{u}_x)=0
\end{equation}
\begin{equation}
\rho_e(-\text{i}\bar{\omega} \hat{u}_z)=-\dfrac{d\hat{P}}{d\bar{z}}-\hat{\rho}\left(\bar{z}+\dfrac{d\Phi_e}{d\bar{z}}\right)-\rho_e\dfrac{d\hat{\Phi}}{d\bar{z}}
\end{equation}
\begin{equation}
\left(\dfrac{d^2}{d\bar{z}^2}-\bar{k}_x^2\right) \hat{\Phi}=\dfrac{\hat{\rho}\Delta}{Q_{2D_0}}
\end{equation}
Solutions to this problem without self-gravity  have been produced
by \citet{lubow93} for an isothermal stratified disc and
by \citet{kory95} and \citet{ogilvie98waves} 
in the case of a polytropic disc. More recently the
self-gravitating problem was tackled by \citet{mamat10}.

The eigenfunctions in an isothermal non-self-gravitating disk are the Hermite polynomials
and the dispersion relation is
\begin{equation}
\label{eq_disp_vertical}
(-\bar{\omega}^2+n)(-\bar{\omega}^2+1)-\bar{k}_x^2 \bar{\omega}^2=0
\end{equation}
where $n$ is the order of the Hermite polynomial and is
related to the vertical structure of the mode (the number
of nodes in the eigenfunctions).
There exists two branches of solutions for each $n>0$, a low frequency
branch with $\bar{\omega} <1$ which has an inertial character and a
high frequency branch with $\bar{\omega} >1$ which has an acoustic
character. Note that the case
$n=0$ is often considered separately. At frequencies larger than the
epicyclic, it corresponds to a two-dimensional
acoustic-inertial wave, which can be associated with an f-mode (free
surface oscillation) for the polytropic case. 

\begin{figure}
\centering
\includegraphics[width=\columnwidth]{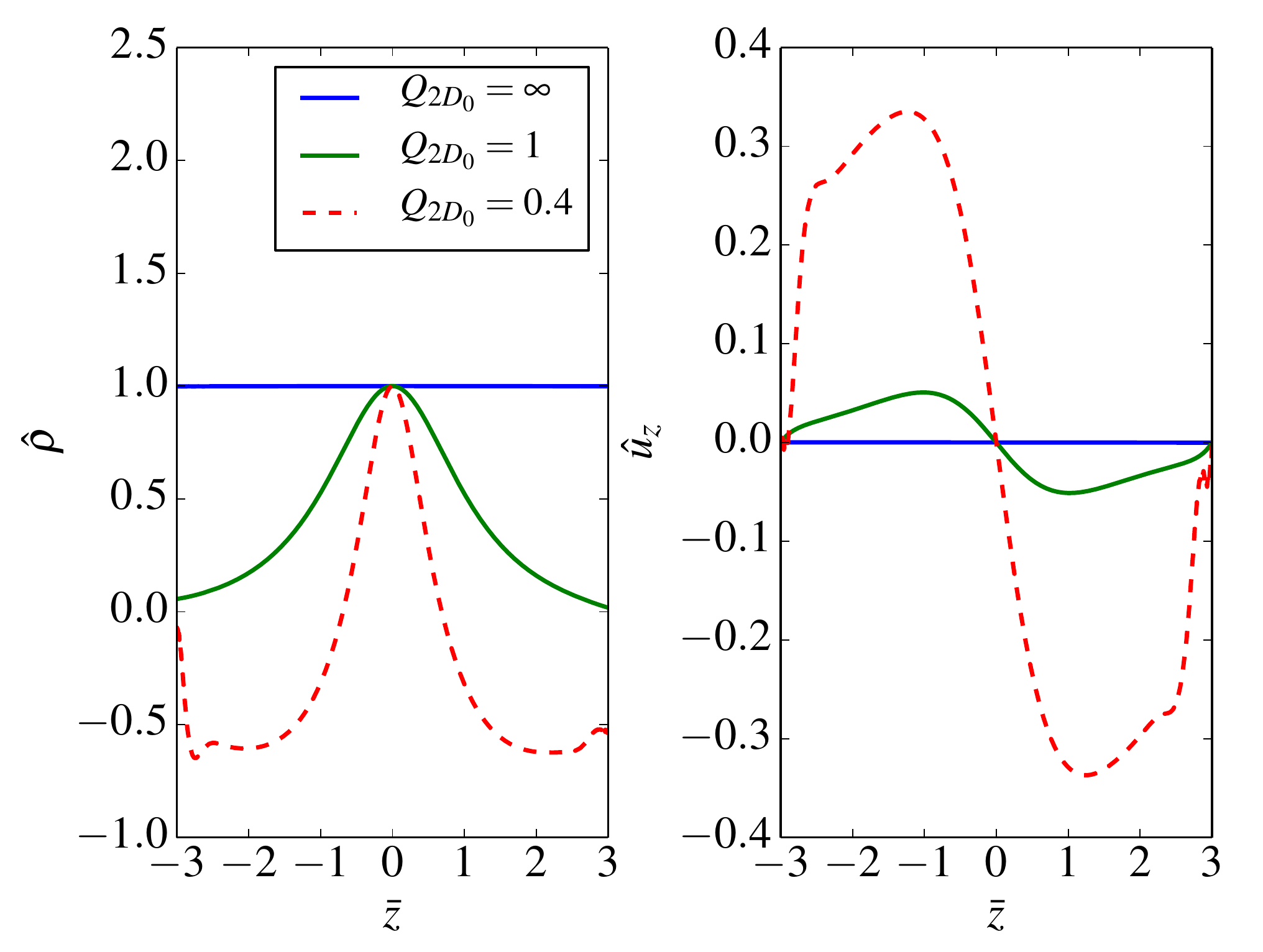} 
\caption{Vertical structure of the $n=0$ eigenmode for 3 different $Q_{{2D}_{0}}=\{\infty, \,\, 1, \,\, 0.4\}$. Left to right: density and vertical velocity perturbations. Note that the mode is normalized such that $\hat{\rho}(z=0)=1$. The noise at the boundary (in particular for $Q_{{2D}_{0}}=0.4$) results from the fact that we are solving the linear problem for the momentum and not for the velocity components directly.}
\label{fig_eigenfunction}
\end{figure}

\begin{figure}
\centering
\includegraphics[width=0.9\columnwidth]{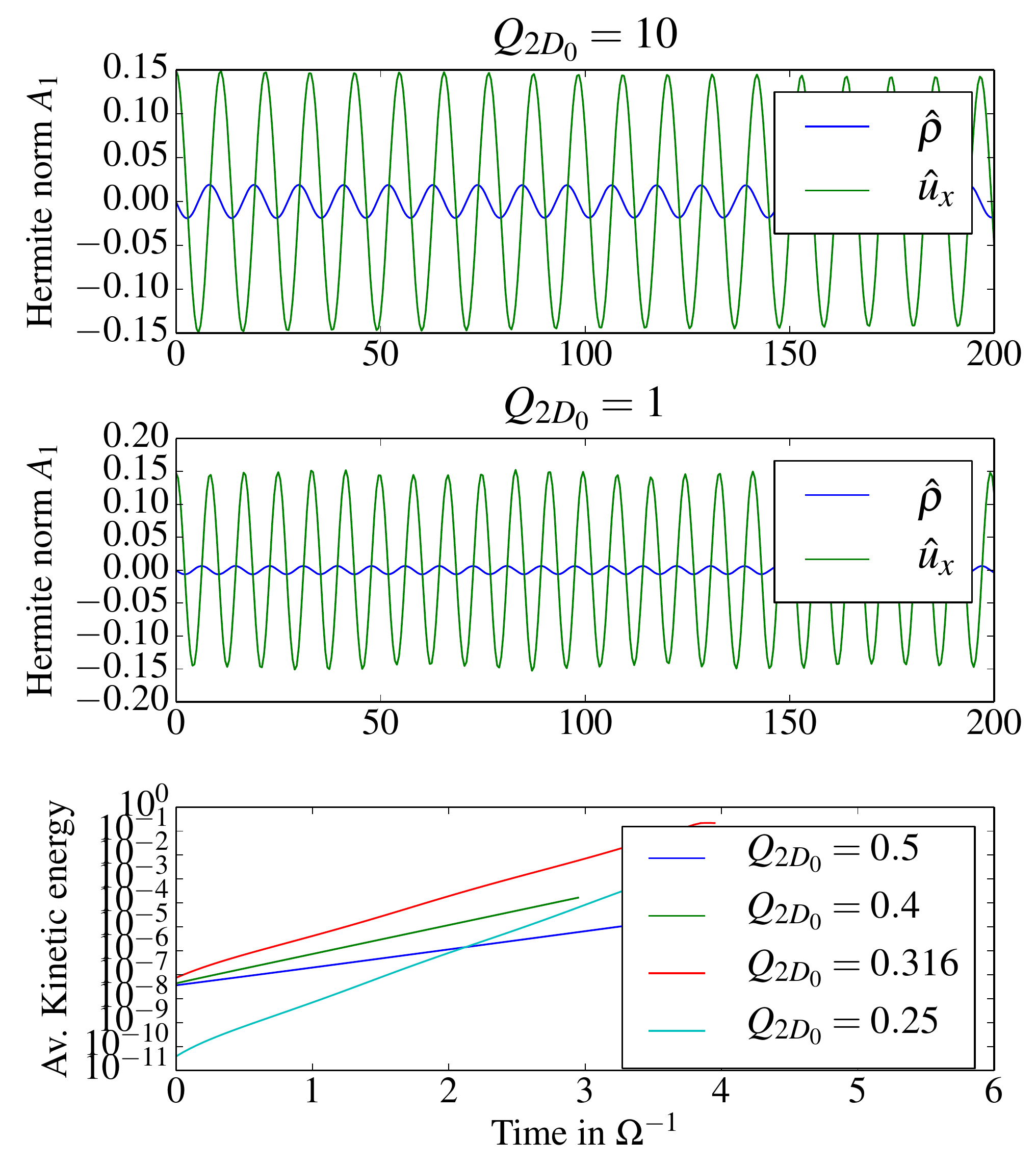} 
\caption{The top two panels describe the time evolution of the
  density and radial velocity perturbation of the $n=1$ inertial mode
  for different values of $Q_{{2D}_{0}}$ (amplitudes of perturbations
  are expressed in terms of the Hermite norm, defined through Eq.~\eqref{Hnorm}) 
The last panel shows the evolution of the average kinetic energy for the $n=0$ mode for four values of $Q_{{2D}_{0}}$. }
\label{fig_linearmode}
\end{figure}

We restrict our study to the isothermal case and
a fixed $\bar{k}_x=2\pi/5$. The surface density is set to
$\Sigma=1$. For a given value of the Toomre parameter $Q_{{2D}_{0}}$,
we first compute the corresponding background density and
gravitational potential, using the methods described in Appendix \ref{appA} (the
equilibria 
fixes the value of $\Delta$, which depends on $Q_{{2D}_{0}}$).
We then compute the eigenvalues  $\bar{\omega}_i$ and corresponding
eigenmodes of the linearized system, using a Chebyshev collocation
method on a Gauss-Lobatto grid. This method results in a matrix
eigenvalue problem that can be solved using a QZ method
\citep{golub96,boyd01}. Numerical convergence is guaranteed by
comparing eigenvalues for different grid resolutions and eliminating
the  spurious ones. The vertical domain has a size $6H_0$ and is
resolved by a maximum resolution of 600 points. Boundary conditions
for perturbed modes are $\rho_e \mathbf{\hat{u}}=0$ (momentum tends to
0) and $d\hat{\Phi}/d\hat{z}=\pm \hat{k_x}\hat{\Phi}$ 
 (justified if the background density decreases quasi-exponentially).\\

When self-gravity is negligible, $Q_{{2D}_{0}}=\infty$, we checked that the eigenvalues
match the dispersion relation (\ref{eq_disp_vertical}) and that for
$n>0$ there exist two different types of solutions corresponding to
either $\bar{\omega}>1$ (p modes) or $\bar{\omega}<1$ (r modes). We
then vary $Q_{{2D}_{0}}$ and focus first on the modes with inertial
character. For each branch labelled $n=1,2,...,7$, ($n$ defined as the
order of the Hermite polynomial in the limit $Q_{{2D}_{0}}=\infty$),
we plot in the top panel of Fig.~\ref{fig_linearstab} the evolution of the wave
frequency as a function of $Q_{{2D}_{0}}$. We find that $\bar{\omega}$
decreases with increasing $Q_{{2D}_{0}}$ but never becomes complex (unstable), at
least for the range of $Q_{{2D}_{0}}$ studied
($0.25<Q_{{2D}_{0}}<\infty$). Instead, the gravitational unstable mode
is associated with the inertial-acoustic branch $n=0$ (f modes). The
corresponding eigenfunction has no structure in $z$ for
$Q_{{2D}_{0}}=\infty$ but appears to develop some structure as
$Q_{{2D}_{0}}$ is decreased. Figure \ref{fig_eigenfunction} shows in
particular that the  vertical velocity component of the unstable mode
develops an odd symmetry as for $n=1$.

Having solved the linear problem directly, we check that our version
of PLUTO reproduces its main results 
in simulations of stable and unstable axisymmetric
waves. For that purpose, we used a box of size $5H \times 5H \times 6H$
(full disc in $z$) and introduced at $t=0$ a perturbation with $k_x=2
\pi/L_x$, $k_y=0$ whose vertical shape is determined by the
eigensolver (in addition to the background density
equilibrium). Figure \ref{fig_linearmode} shows the density and radial
velocity amplitude of the $n=1$ inertial mode, expressed in terms of
the Fourier-Hermite norm
\begin{equation}\label{Hnorm}
A_n (X)=\int\int \hat{X}(z) \cos(k_xx)H_{e_n}(z)\mathrm{e}^{-z^2/2} d_x d_z
\end{equation}
for $Q_{{2D}_{0}}=1$ and $10$. At a resolution
of $256 \times 1 \times 256$, the solution remains periodic and
conserves its vertical shape for more than $200\,\Omega^{-1}$. The
wave frequencies, inferred from these plots, are superimposed on the
dispersion diagram of Fig.~\ref{fig_linearstab} and compared with the
theoretical frequencies,  for different $Q_{{2D}_{0}}$  (red stars
denoted a resolution of $256 \times 1 \times 256$ whereas green stars
is for a resolution of $512 \times 1 \times 512$). We find a very good
agreement between the frequencies computed with the eigensolver and
those measured from the simulations.  The relative error remains
always smaller than 1\% (typically the mean error is 0.3\%). We have
undertaken 
a similar test for the $n=2$ branch, although only 2 points have
been computed in that case. 

Finally we simulated the
$n=0$ branch which becomes unstable for $Q_{{2D}_{0}}=0.63$. Fig
\ref{fig_linearmode} (bottom) shows the evolution of the averaged
kinetic energy in the box for different $Q_{{2D}_{0}}$. Growth rates
are easily measured by computing the slope of these curves. In Fig
\ref{fig_linearstab} (bottom panel), we superimpose the numerical
growth rate obtained (green stars) with the predicted  
growth rate as a function of $Q_{{2D}_{0}}$. We found again that
numerical 
and theoretical growth rate match remarkably well, even for small values of $Q_{{2D}_{0}}$.

\section{Nonlinear `instability' of axisymmetric modes}

\begin{figure}
\centering
\includegraphics[width=\columnwidth]{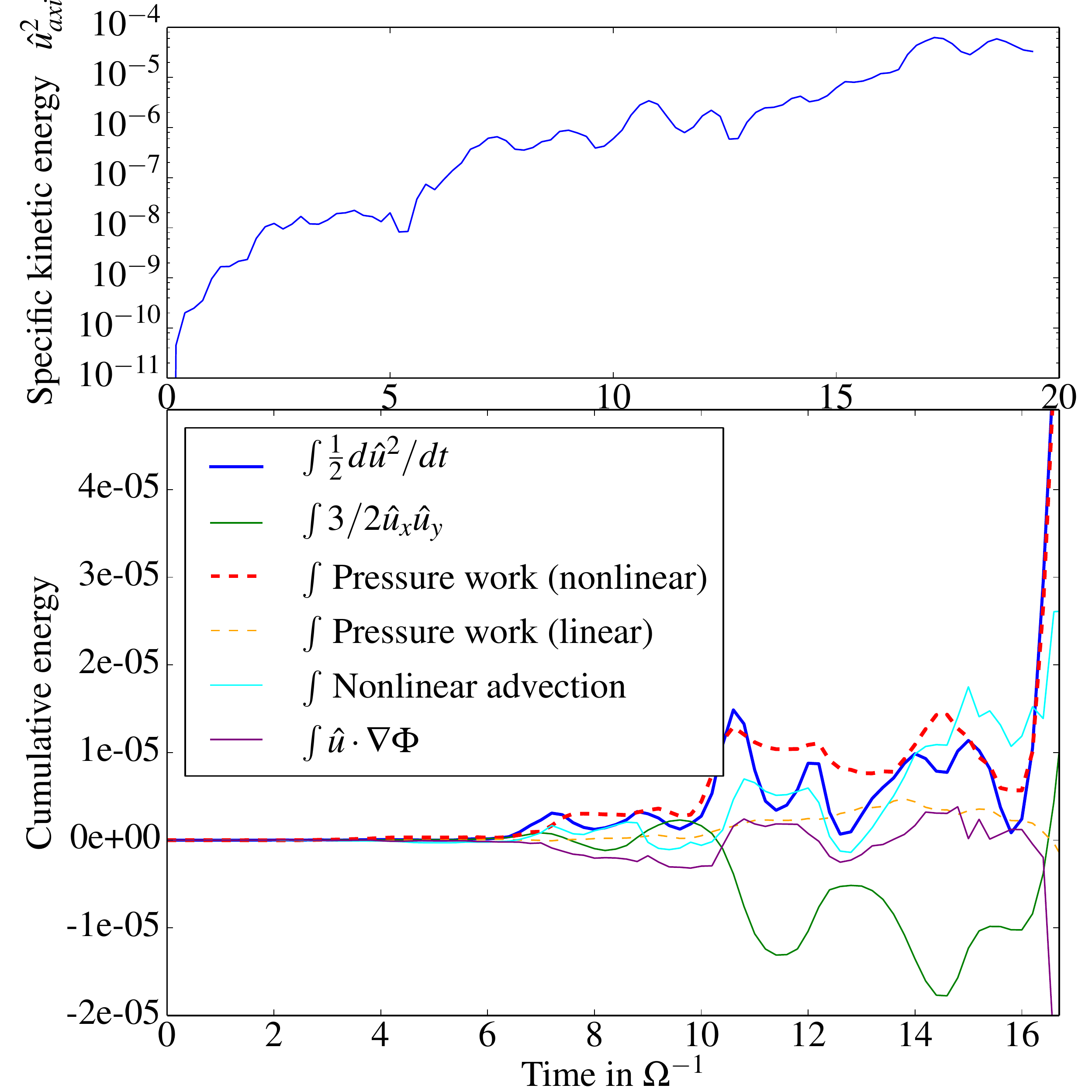} 
\caption{Top: Time evolution of the specific kinetic energy in the
  midplane $z=0$ of the fundamental axisymmetric mode $k_x=k_{x_0}$ in
  the box $L_x=40 H_0$. Bottom: energy budget of the fundamental
  axisymmetric perturbation. Each curve represents the cumulative work
  associated with
  a term (or a force) in the Navier-Stockes equation.}
\label{fig_axi_growth}
\end{figure}

Simulations of gravito-turbulence in shear flows exhibit strong large
scale axisymmetric motions (see Section \ref{sec_fourier}). To better
understand how these motions are generated by the turbulent flow, we
perform a simple experiment: we considered a gravito-turbulent state
in a small box $L_x=20 H_0$ and we used the symmetry of the shearing
box to construct a stacked version of this state in a box of twice the
original size. We then examined the midplane time-evolution of the fundamental
axisymmetric mode $k_x=k_{x_0}$, which possesses a wavelength the size
of the new box. The results are summarized
in Fig.~\ref{fig_axi_growth}. At $t=0$, the energy associated with this
mode is zero, since we started from a state computed in box of half
size. The mode is forced by non-linearities and grows with
a typical growth rate $\gamma_{a}=0.3\Omega$. Its short term  
evolution is very stochastic, but its long-term evolution is more or
less exponential. 

To understand the nature  of this instability, we investigated the kinetic energy budget of this fundamental mode: 
\begin{equation}
\dfrac{1}{2}\dfrac{d  \mathbf{u_F}^2}{dt}=\frac{3}{2}u_{x_F} u_{y_F}+\mathbf{u_F} \cdot \nabla \mathbf{\Phi} -\mathbf{u_F} \cdot \left(\dfrac{\nabla P}{\rho}\right)_{F}-\mathbf{u_F} \cdot (\mathbf{u} \cdot \nabla \mathbf{u})_F
\end{equation}
The subscript `F' denotes here the fundamental axisymmetric mode $k_x=k_{x_0}$ or its related projection for nonlinear terms . On the
right hand side, the two first linear terms are associated with the background shear and self-gravity. The third and fourth terms are related to the pressure gradient and nonlinear advection. The pressure gradient term can be decomposed into a linear and nonlinear part: 
\begin{equation}
\left(\dfrac{\nabla P}{\rho}\right)_{F}=\left(\dfrac{\nabla P}{\rho}\right)^{NL}_{F}+\left(\dfrac{\nabla {P}_{F}}{\rho_0}\right)
\end{equation}
where $\rho_0$ is the midplane background density. 

Figure \ref{fig_axi_growth} (bottom) shows the cumulative contribution
of each term involved in the mode growth. First, we found that the work
done by self-gravity is mainly negative and hence does not participate in the
instability. This is expected since the axisymmetric mode is
gravitationally linearly stable. The shear term $-Su_{x_0} u_{y_0}$ is
also stabilizing. Note that the Coriolis force does no work since it
only redistributes energy from the $y$ component to the $x$ component
or vice-versa.

 The main contribution to the kinetic energy, at least for the first orbits, is clearly the nonlinear term associated with the pressure work. 
Physically, this term corresponds to a baroclinic effect, due to the misalignement of pressure and density gradients. Indeed, if the pressure term is assumed to be dominant in the energy evolution of the axisymmetric mode, then the vorticity equation reduces to:
\begin{equation}
\dfrac{\partial \bm{\omega}}{\partial t}=\dfrac{\bm{\nabla} P \times \bm{\nabla} \rho}{\rho^2}.
\end{equation}
The formation of axisymmetric cells associated with vorticity aligned
in the $y$ direction is possible if baroclinic non-axisymmetric
perturbations (or spiral waves) interact with each other. In other
words, if the average in $y$ of $({\bm{\nabla} P \times \bm{\nabla}
  \rho})_F \cdot \bm{\hat{y}}$  has a positive feedback. 

A possible origin for the global process is a triadic interaction
between the $m=1$ spiral density waves and axisymmetric modes.  The
scenario can be described as follows: non-axisymmetric waves tap into
two reservoirs of energy which are the shear and the self-gravity. The
nonlinear interaction between a swinging leading wave and a dying
trailing wave produces an axisymmetric modulation (through the
baroclinic term), which in turn reinforces the non-axisymmetric waves,
via a mechanism similar to that described in \cite{lithwick07}. This scenario  
provides a potential explanation for why axisymmetric modes are
growing exponentially, 
despite the absence of a classical `linear' instability.

\label{appc}
\label{lastpage}
\end{document}